\documentclass{aa} 

\usepackage{hyperref}
\usepackage{graphicx}
\usepackage{txfonts}
\usepackage{xspace}   % neat package to handle spaces in \newcommands
\newcommand{\HI}{H\,\textsc{i}\xspace}
\newcommand{\Msol}{M$_{\odot}$\xspace}
\newcommand{\Mstar}{M$_{\star}$\xspace}
\newcommand{\kms}{~km\,s$^{-1}$\xspace}

\newcommand{\FHI}{F$_{\rm HI}$\xspace}
\newcommand{\FHmol}{F$_{\rm H2}$\xspace}
\newcommand{\Ha}{H$\alpha$\xspace}

\newcommand{\SoFiA}{\texttt{SoFiA}\xspace}
\newcommand{\CARACal}{\texttt{CARACal}\xspace}

\begin{document} 

    \title{A MeerKAT view of pre-processing in the Fornax A group}

    \titlerunning{Pre-processing in the Fornax A group}
  
   \author{D.~Kleiner\inst{1}, P.~Serra\inst{1}, F.~M.~Maccagni\inst{1}, A.~Venhola\inst{2}, K.~Morokuma-Matsui\inst{3},  R.~Peletier\inst{4}, E.~Iodice\inst{5}, M.~A.~Raj\inst{5}, W.~J.~G.~de Blok\inst{6,7,4}, A.~Comrie\inst{8}, G.~I.~G.~J\'ozsa\inst{9,10,11}, P.~Kamphuis\inst{12}, A.~Loni\inst{1,13}, S.~I.~Loubser\inst{14}, D.~Cs.~Moln\'ar\inst{1}, S.~S.~Passmoor\inst{9}, M.~Ramatsoku\inst{10,1}, A.~Sivitilli\inst{7}, O.~Smirnov\inst{10,9}, K.~Thorat\inst{15,8} \and F.~Vitello\inst{16}  
          }
   \authorrunning{Kleiner et al.}

   \institute{INAF – Osservatorio Astronomico di Cagliari, Via della Scienza 5, 09047 Selargius, CA, Italy\\
              \email{dane.kleiner@inaf.it}
         \and
         University of Oulu, Space physics and astronomy unit, Pentti Kaiteran katu 1,  90014, Oulu, Finland
         \and
         Institute of Astronomy, Graduate School of Science, The University of Tokyo, 2–21–1 Osawa, Mitaka, Tokyo 181–0015, Japan
         \and
         Kapteyn Astronomical Institute, University of Groningen,PO Box 800, 9700 AV Groningen, The Netherlands
         \and
         INAF -- Astronomical observatory of Capodimonte, Via Moiariello 16, Naples 80131, Italy
         \and
         Netherlands Institute for Radio Astronomy (ASTRON), Oude Hoogeveensedijk 4, 7991 PD Dwingeloo, the Netherlands
         \and
         Deptarment of Astronomy, Univ. of Cape Town, Private Bag X3, Rondebosch 7701, South Africa
         \and
         Inter-University Institute for Data Intensive Astronomy, University of Cape Town, Cape Town, Western Cape, 7700, South Africa
         \and
         South African Radio Astronomy Observatory, 2 Fir Street, Black River Park, Observatory, Cape Town, 7925, South Africa
         \and
         Department of Physics and Electronics, Rhodes University, PO Box 94, Makhanda, 6140, South Africa
         \and
         Argelander-Institut f\"ur Astronomie, Auf dem H\"ugel 71, D-53121 Bonn, Germany
         \and
         Ruhr University Bochum, Faculty of Physics and Astronomy, Astronomical Institute, 44780 Bochum, Germany
         \and
         Dipartimento di Fisica, Universit\`{a} di Cagliari, Cittadella Universitaria, 09042 Monserrato, Italy
         \and
         Centre for Space Research, North-West University, Potchefstroom 2520, South Africa
         \and
         Department of Physics, University of Pretoria, Private Bag X20, Hatfield 0028, South Africa
         \and
         INAF -- Istituto di Radioastronomia, via Gobetti 101, I-40129 Bologna, Italy\\
             }

   \date{Received 13 November, 2020; accepted 25 January, 2021}
 
  \abstract{We present MeerKAT neutral hydrogen (\HI) observations of the Fornax\,A group, which is likely falling into the Fornax cluster for the first time. Our \HI image is sensitive to 1.4 $\times$ 10$^{19}$ atoms cm$^{-2}$ over 44.1\kms, where we detect \HI in 10 galaxies and a total of (1.12 $\pm$ 0.02) $\times$ 10$^{9}$ \Msol of \HI in the intra-group medium (IGM). We search for signs of pre-processing in the 12 group galaxies with confirmed optical redshifts that reside within the sensitivity limit of our \HI image. There are 9 galaxies that show evidence of pre-processing and we classify each galaxy into their respective pre-processing category, according to their \HI morphology and gas (atomic and molecular) scaling relations. Galaxies that have not yet experienced pre-processing have extended \HI discs and a high \HI content with a H$_2$-to-\HI ratio that is an order of magnitude lower than the median for their stellar mass. Galaxies that are currently being pre-processed display \HI tails, truncated \HI discs with typical gas fractions, and H$_2$-to-\HI ratios. Galaxies in the advanced stages of pre-processing are the most \HI deficient. If there is any \HI, they have lost their outer \HI disc and efficiently converted their \HI to H$_2$, resulting in H$_2$-to-\HI ratios that are an order of magnitude higher than the median for their stellar mass. The central, massive galaxy in our group (NGC\,1316) underwent a 10:1 merger $\sim$\ 2\,Gyr ago and ejected 6.6 -- 11.2 $\times$ 10$^{8}$ \Msol of \HI , which we detect as clouds and streams in the IGM, some of which form coherent structures up to $\sim$ 220\,kpc in length. We also detect giant ($\sim$ 100\,kpc) ionised hydrogen (\Ha) filaments in the IGM, likely from cool gas being removed (and subsequently ionised) from an in-falling satellite. The \Ha filaments are situated within the hot halo of NGC\,1316 and there are localised regions that contain \HI. We speculate that the \Ha and multiphase gas is supported by magnetic pressure (possibly assisted by the NGC\,1316 AGN), such that the hot gas can condense and form \HI that survives in the hot halo for cosmological timescales.}


   \maketitle
%
%-------------------------------------------------------------------

\section{Introduction}
Our current understanding of galaxy formation and evolution is that secular processes and galaxy environment fundamentally shape the properties of galaxies \citep[e.g.][]{Baldry2004, Balogh2004, Bell2004, Peng2010, Driver2011, Schawinski2014, Davies2019}. In the local Universe (z $\sim$ 0) up to $\sim$ 50\% of galaxies reside in groups \citep{Eke2004, Robotham2011}, making it essential to understand the group environment in the context of galaxy evolution.

While there is no precise definition of a galaxy group, it generally contains 3 -- 10$^{2}$ galaxies in a dark matter (DM) halo of 10$^{12}$ -- 10$^{14}$ \Msol \citep[e.g.][]{Catinella2013}. As the galaxy number density and DM halo mass of groups span a wide range, there is no dominant transformation mechanism that galaxies are subjected to, but rather multiple secular and external mechanisms working together. The properties of group galaxies appear to correlate with group halo mass and virial radius, implying that quenching paths in groups are different from those in clusters \citep{Weinmann2006, Haines2007, Wetzel2012, Woo2013, Haines2015}. 

As galaxies fall towards clusters, there is sufficient time for external (i.e. environmentally driven, such as tidal and hydro-dynamical) mechanisms to transform and even quench the galaxies, prior to reaching the cluster \citep[e.g.][]{Porter2008, Haines2013, Haines2015, Bianconi2018, Fossati2019, Ruchika2020}. This is called ``pre-processing'' and refers to the accelerated, non-secular evolution of galaxies that occurs prior to entering a cluster. As pre-processing requires external mechanisms to transform the galaxies, this evolution commonly occurs in groups, where it is generally thought that group galaxies follow a different evolutionary path compared to galaxies of the same mass in the field \citep[e.g.][]{Fujita2004, Mahajan2013, Roberts2017, Cluver2020}. In particular, pre-processing is likely to be most efficient in massive ($>$ 10$^{10.5}$ \Msol) galaxies residing in massive (10$^{13}$ -- 10$^{14}$ \Msol) groups \citep{Donnari2020}. It has also been shown that pre-processing is responsible for the decrease in star formation activity for late-type galaxies at distances between 1 and 3 cluster virial radii \citep[e.g][]{Lewis2002, Gomez2003, Verdugo2008, Mahajan2012, Haines2015}.

Neutral hydrogen in the atomic form (\HI) is ideal for tracing tidal and hydro-dynamical processes in galaxies and the intra-group medium (IGM). \HI is the main component of the interstellar medium (ISM) and can show the effects of ram pressure, viscous and turbulent stripping, thermal heating \citep[e.g.][]{Cowie1977, Nulsen1982, Chung2007, Rasmussen2008, Chung2009, Steinhauser2016, Ramatsoku2020}, and moderate and strong tidal interactions \citep[e.g.][]{Koribalski2012, deBlok2018, Kleiner2019}, long before these mechanism can be identified in the stars.

In this paper we present a detailed analysis of the Fornax\,A galaxy group based on \HI and ancillary observations. The Fornax\,A group is an excellent candidate to search for pre-processing signatures as it is likely in-falling into the (low mass -- 5 $\times$ 10$^{13}$ \Msol) Fornax cluster \citep{Drinkwater2001} for the first time. The group galaxies span a variety of stellar masses and morphological types, implying that tidal and hydro-dynamical interactions are likely to affect the galaxies gas and stellar content \citep{Raj2020}. 

Using Meer Karoo Array Telescope (MeerKAT) \HI observations, deep optical imaging from the Fornax Deep Survey \citep[FDS:][]{Iodice2016, Iodice2017, Venhola2018, Venhola2019, Raj2019, Raj2020}, wide-field \Ha imaging from the VLT Survey Telescope (VST) and molecular gas observations from the Atacama Large Millimetre Array (ALMA), we identify galaxies at different stages of pre-processing following various types of interactions.

This paper is organised as follows: Section \ref{sec:group_desc} describes the Fornax\,A group. Section \ref{sec:obs} describes the \HI and \Ha observations and the data reduction process used to produce our images. We present the results of our \HI measurements, \HI images, and the relation to stellar and \Ha emission in Section \ref{sec:results}. In Section \ref{sec:discussion} we present the atomic-to-molecular gas ratios and discuss the evidence and timescale of pre-processing in the group. Finally, we summarise our results in Section \ref{sec:conclusions}. Throughout this paper we assume a luminosity distance of 20\,Mpc to the most massive galaxy (NGC\,1316) in the Fornax\,A group \citep{Cantiello2013, Hatt2018} and assume all objects in the group are at the same distance. At this distance, 1\arcmin\xspace corresponds to 5.8\,kpc.

%--------------------------------------------------------------------

\section{The Fornax A group}
\label{sec:group_desc} 
The Fornax\,A galaxy group is the brightest group in the Fornax volume. It is located on the cluster outskirts at a projected distance of $\sim$ 3.6 deg (1.3\,Mpc, or $\sim$ 2 $\times$ the Fornax cluster virial radius) from the cluster centre and has a mass of 1.6 $\times$ 10$^{13}$ \Msol, which is of the same order of magnitude as the Fornax cluster (M$_{\rm vir}$ $\sim$ 5 $\times$ 10$^{13}$ \Msol) itself \citep{Maddox2019}. Within the virial radius of the group \citep[$\sim$ 1 degree or 0.38\,Mpc, as measured by][]{Drinkwater2001}, there are approximately 70 galaxies (mostly dwarfs) that have been photometricially identified as likely group members \citep{Venhola2019}, of which 13 have confirmed spectroscopic redshifts \citep{Maddox2019}.

The brightest group galaxy (BGG), NGC\,1316, is a peculiar early-type galaxy with a stellar mass of 6 $\times$ 10$^{11}$ \Msol \citep{Iodice2017}. NGC\,1316 is a giant radio galaxy \citep{Ekers1983, Formalont1989, McKinley2015, Maccagni2020}, known merger remnant, and the brightest galaxy in the Fornax cluster volume (even brighter than the brightest cluster galaxy NGC\,1399). There are a number of extended stellar loops and streams in NGC\,1316 that are a result of a 10:1 merger that occurred 1 -- 3\,Gyr ago, between a massive early-type galaxy and a gas-rich, late-type galaxy \citep{Schweizer1980, Mackie1998, Goudfrooij2001, Iodice2017, Serra2019}. The majority of the remaining bright ($m_{B}$ < 16) galaxies are late types that have  stellar mass ranges of 8 $<$ log(\Mstar /\Msol) $<$ 10.5 \citep{Raj2020}. 

There have been a variety of previous studies that have detected \HI in the Fornax\,A group. \citet{Horellou2001} and \citet{Serra2019} imaged the central region of the Fornax\,A group in \HI, where the more recent image of \citet{Serra2019} detected NGC\,1316, NGC\,1317, NGC\,1310, and ESO\,301-IG\,11, along with four clouds at the outskirts of NGC\,1316 (EELR, SH2, C$_{\rm N, 1}$, and C$_{\rm N, 2}$), and two tails (T$_{\rm N}$ and T$_{\rm S}$). The remaining six galaxies, which  have previously been detected, are NGC\,1326, NGC\,1326A ,and NGC\,1326B in the \HI Parkes All Sky Survey  \citep[HIPASS;][]{Meyer2004, Koribalski2004b}, NGC\,1316C with the Nan\c{c}ay telescope \citep{Theureau1998}, FCC\,35 with the Australian Telescope Compact Array (ATCA) and the Green Bank Telescope \citep{Putman1998, Courtois2015}, and FCC\,46 with the ATCA \citep{deRijcke2013}. Within NGC\,1316, \HI has been resolved in the centre and correlates with massive amounts of molecular gas \citep{Morokuma-Matsui2019, Serra2019}. \HI has also been detected in the outer stellar halo, within the regions defined by the \Ha extended emission line region \citep[EELR; originally discovered by][]{Mackie1998}, in the southern star cluster complex \citep[SH2;][]{Horellou2001} and in two northern clouds (C$_{\rm N,1}$ and C$_{\rm N,2}$) \citep{Serra2019}. Lastly, $\sim$ 6 $\times$ 10$^{8}$ \Msol of \HI was detected in the IGM, defined as the northern and southern tails (T$_{\rm N}$ and T$_{\rm S}$). The tails are ejected \HI gas from the NGC\,1316 merger and extend up to 150\,kpc from the galaxy centre \citep{Serra2019}.

The Fornax\,A group is an ideal system to search for pre-processing. Evidence suggests that the group is in the early stage of assembly \citep{Iodice2017, Raj2020} and is located at the cluster infall distance where pre-processing is thought to occur \citep{Lewis2002, Gomez2003, Verdugo2008, Mahajan2012, Haines2015}. The BGG is massive enough to experience efficient pre-processing \citep{Donnari2020} and \citet{Raj2020} show that there are signatures of pre-processing in the group; six of the nine late types have an up-bending (type III) break in their radial light profile. This indicates that the star formation may be halting in the outer disc of galaxies, although, what is driving the decline in star formation is not yet clear. 

%--------------------------------------------------------------------

\section{Observations and data reduction}
\label{sec:obs}

\subsection{MeerKAT radio observation}
Commissioned in July 2018, MeerKAT is a new radio interferometer and a precursor for the Square Kilometre Array SKA1-MID telescope \citep{Jonas2016, Mauch2020}. MeerKAT is designed to produce highly sensitive radio continuum and \HI images with good spatial and spectral resolution in a relatively short amount of observing time. The MeerKAT Fornax Survey (MFS; PI: P.Serra) is one of the designated Large Survey Projects (LSPs) of the MeerKAT telescope. The MFS will observe the Fornax galaxy cluster in \HI over a wide range of environment densities, down to a column density of a few $\times$ 10$^{19}$ atoms cm$^{-2}$ at a resolution of 1\,kpc, equivalent to a \HI mass limit of 5 $\times$ 10$^{5}$ \Msol \citep{Serra2016}. 

The Fornax\,A group was observed with MeerKAT in two different commissioning observations in June 2018, which differ by the number of antennas (36 and 62, respectively) connected to the correlator. We present the details of these observations and of the \HI cube in Table \ref{tab:HI_obs}. The MeerKAT baselines range between 29\,m and 7.7\,km and for both these observations, the SKARAB correlator in the 4k mode was used, which consists of 4096 channels in full polarisation in the frequency range 856-1712\,MHz with a resolution of 209\,kHz (equivalent to 44.1\kms for \HI at the distance of the Fornax cluster).

The first observation (referred to as Mk-36) used 36 antennas and observed the target for a total of 8  h. Results from this observation are presented both in radio continuum \citep{Maccagni2020} and in \HI \citep{Serra2019}; these papers provide a detailed description of the data reduction process. In this work, we use the Mk-36 calibrated measurement set in combination with that from the second observation (detailed below). 

The second observation (Mk-62) used 62 antennas and observed the target for a total of 7 h. PKS 1934-638 and PKS 0032-403 were observed, where the former was observed for 20 min and used as the bandpass and flux calibrator while the latter was observed for 2 min every 10 min and used as the gain calibrator. 

% Table of basic observation properties
\begin{table*}
  \begin{center}
    \caption{Observation and \HI cube properties. The measurements of the \HI cube RMS noise and column density (over a single channel of 44.1\kms) were taken in the pointing centre and restoring beam was taken from the centre channel.}
    \label{tab:HI_obs}
    \begin{tabular}{c c c}
    
        \hline
        \hline
        Property & Mk-36 observation & Mk-62 observation\\
        \hline        
        Date & 2 June 2018 & 16 June 2018\\
        ID   & 20180601-0009  & 20180615-0039\\
        Time on target  & 8 hr & 7 hr\\
        Number of antennas & 36 & 62\\
        Pointing centre (J2000) &  \multicolumn{2}{c}{03h 22m 41.7s, -37d 12\arcmin\xspace 30.0\arcsec}\\
        Available bandwidth  & \multicolumn{2}{c}{856 - 1712 MHz}\\
%        Continuum frequency range & 1330 - 1450 MHz\\
        \HI cube frequency range  & \multicolumn{2}{c}{1402 - 1420 MHz}\\
        \HI cube spectral resolution & \multicolumn{2}{c}{209 kHz (44.1\kms at z $\sim$ 0)}\\
        \HI cube pixel size & \multicolumn{2}{c}{6.5\arcsec}\\
        \HI cube weight & \multicolumn{2}{c}{robust = 0.5 and 20\arcsec taper} \\
        \HI cube RMS noise & \multicolumn{2}{c}{90 $\mu$Jy beam$^{-1}$}\\
        \HI cube restoring beam & \multicolumn{2}{c}{33.0\arcsec $\times$ 29.2\arcsec} \\ % @ 120 deg PA
        3$\sigma$ \HI column density & \multicolumn{2}{c}{1.4 $\times$ 10$^{19}$ atoms cm$^{-2}$}\\

        \hline
    \end{tabular}
  \end{center}
\end{table*}

We used the Containerised Automated Radio Astronomical Calibration \citep[\texttt{CARACal}\footnote{\url{https://caracal.readthedocs.io}}; ][]{Jozsa2020} pipeline to reduce the MeerKAT observations. The pipeline uses \texttt{Stimela}\footnote{\url{https://github.com/SpheMakh/Stimela}}, which containerises different open-source radio interferometry software in a \texttt{Python} framework. This makes the pipeline both flexible and highly customisable and has been used to reduce MeerKAT and other (e.g. Jansky Very Large Array) interferometric observations \citep[e.g. see][]{Serra2019, Maccagni2020, Ramatsoku2020, Ianjamasimanana2020}. 

We used \texttt{CARACal} to reduce the Mk-62 observation end-to-end and include the already reduced Mk-36 observation \citep{Serra2019, Maccagni2020} at the spectral line imaging step. For the Mk-62 observation, we used 120 (1330 - 1450)\,MHz of bandwidth to ensure adequate continuum imaging and calibration. We used 18 (1402 - 1420)\,MHz, which easily covers the group volume, for the (joint) spectral line imaging. 

Our choice of data reduction techniques and steps is outlined using \CARACal\ as follows: First, we flag the radio frequency interference (RFI) in the calibrators data based on the Stokes Q visibilites using \texttt{AOflagger} \citep{Offringa2012}. Then, a time-independent, antenna-based, complex gain solution was derived for the bandpass using \texttt{CASA bandpass} and the flux scale was determined with \texttt{CASA gaincal}. A Frequency-independent, time-dependent, antenna-based complex gains were determined using \texttt{CASA gaincal}. The gain amplitudes were scaled to bootstrap the flux scale with \texttt{CASA fluxscale}, and the bandpass and complex gain solutions were applied to the target visibilities using \texttt{CASA applycal}. The RFI in the target data was then flagged based on the Stokes Q visibilites, using \texttt{AOflagger} \citep{Offringa2012}. We imaged and self-calibrated the continuum emission of the target with \texttt{WSclean} \citep{Offringa2014, Offringa2017} and \texttt{CUBICAL} \citep{Kenyon2018}, respectively. This process was repeated two more times, in which each self-calibration iteration was frequency-independent and solved only for the gain phase, with a solution interval of 2 min. The final continuum model was subtracted from the visibilities using \texttt{CASA msutils}. The visibilities from both the Mk-36 and Mk-62 calibrated measurement sets were then Doppler corrected into the barycentric rest frame using \texttt{CASA mstransform}. Residual continuum emission in the combined measurement set was removed by fitting and subtracting a second order polynomial to the real and imaginary visibility spectra with \texttt{CASA mstransform}. Then, we created a \HI cube by imaging the \HI emission with \texttt{WSclean} \citep{Offringa2014, Offringa2017} and made a 3D mask through source finding with the Source Findina Application \citep[\SoFiA;][]{Serra2015a}. This was then used as a clean mask to image a new \HI cube with higher image fidelity. Finally, we applied the primary beam correction of \citet{Mauch2020} down to a level of 2\%, which corrects for the sensitivity response pattern of MeerKAT. 

Our \HI cube was imaged\footnote{The deep \HI imaging revealed periodic, artefacts caused by the correlator during this time of commissioning. The artefacts were apparent at the sky position of bright continuum emission. We were able to remove the artefacts by excluding baselines less than 50\,m in the cube and 85\,m for the single, worst channel. While short baselines are essential for diffuse emission, this equates to 5 and 22 baselines out of 1891.} using an 18\,MHz sub-band (centred on NGC\,1316) and the basic properties are presented in Table \ref{tab:HI_obs}. The root mean square (RMS) noise is 90\,$\mu$Jy beam$^{-1}$, which equates to a 3$\sigma$ \HI column density of 1.4 $\times$ 10$^{19}$ atoms cm$^{-2}$ over a single channel of 44.1\kms at the angular resolution of 33.0\arcsec $\times$ 29.2\arcsec. Compared to \citet{Serra2019}, we present an image that is approximately twice as large and more than twice as sensitive\ and has comparable spatial and velocity resolutions.

\begin{figure}

    \includegraphics[width = \columnwidth]{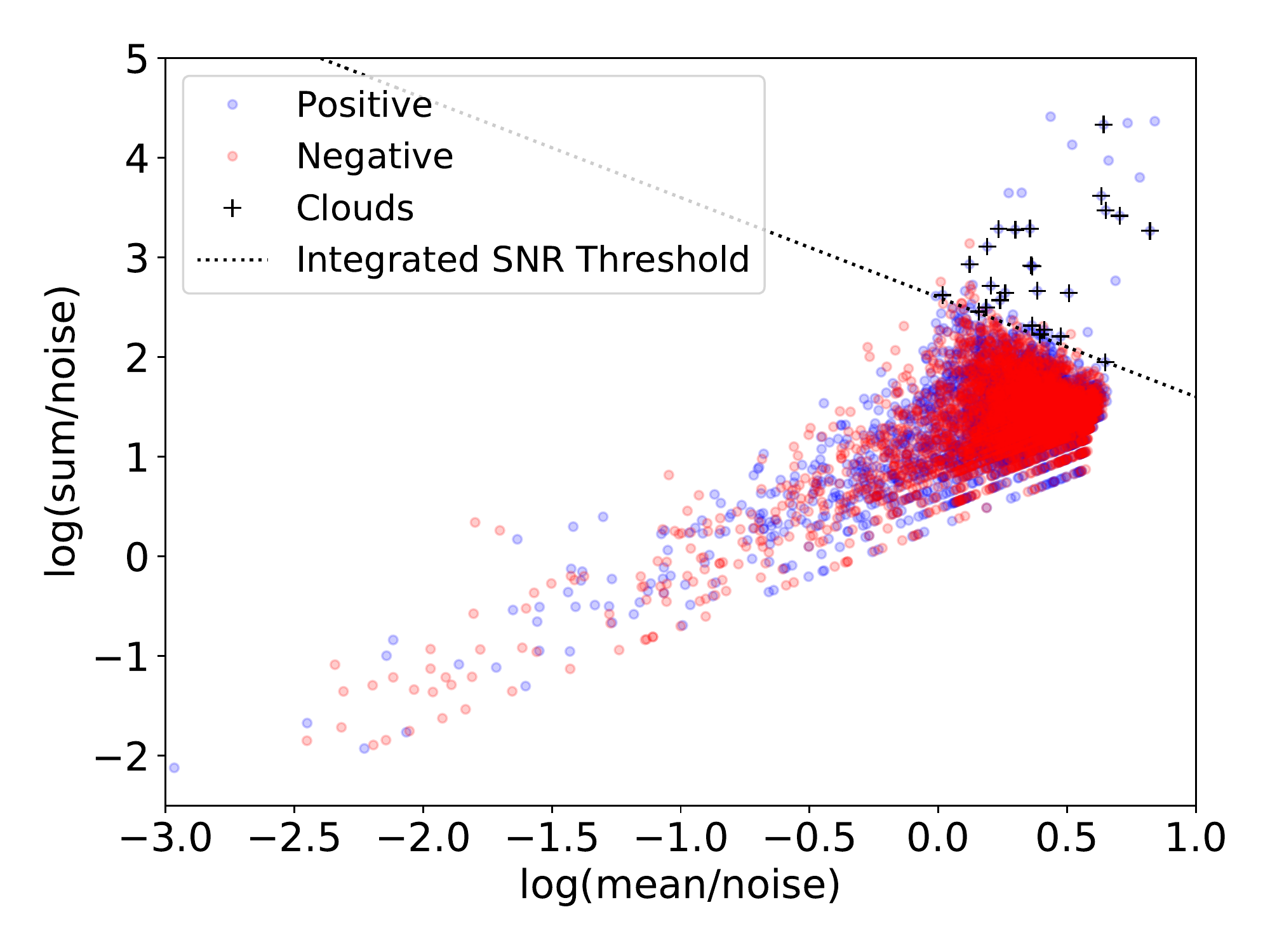}

\caption{Sum of the pixel values as a function of the mean pixel value for all sources detected with \SoFiA. The blue points indicate the positive detections and the red points indicate the negative detections \citep{Serra2012b}. Detected \HI clouds are shown as black crosses. The dotted line shows the per-source integrated S/N of 4. Only positive sources above this threshold and with a reliability $>$ 0.65 are retained in our final catalogue. The chosen integrated S/N of 4 is a conservative threshold as it is closer to area of parameter space occupied by the most statistically significant detections (i.e. the positive sources with a high sum/noise for their mean/noise value) and is clearly above the edge of non-statistically significant detections (i.e. where the density of positive sources is approximately the same as the density of negative sources). Owing to this conservative threshold, the detected \HI clouds, while often diffuse, occupy the parameter space of real, reliable \HI emission.}

\label{fig:detections}
\end{figure}

We searched for \HI sources using \SoFiA outside the \CARACal pipeline. To ensure that we properly captured \HI emission that is diffuse or far from the pointing centre, we tested different combinations of smoothing kernels and detection thresholds in the \SoFiA \texttt{smooth + clip} algorithm, per-source integrated signal-to-noise ratio (S/N) thresholds, and reliability thresholds. Pixels in the \HI cube are detected if their value is above a \texttt{smooth + clip} detection threshold of 3.5 (in absolute value and relative to the cube noise) for spatial smoothing kernels equal to 1, 2, and 3 times the synthesised beam in combination with velocity smoothing kernels over a single (i.e. no smoothing) and three channels. The mean, sum, and maximum pixel value of each detected source (normalised to the local noise) create a parameter space that can separate real \HI emission from noise peaks \citep[Fig. \ref{fig:detections};][]{Serra2012b}. The reliability of each source (defined as the local ratio of positive-to-negative source density within this 3D parameter space) as well as the integrated S/N are then used to identify statistically significant, real \HI sources. Our catalogue was created by retaining only sources with an integrated S/N above 4 and a reliability above 0.65. As shown in Fig. \ref{fig:detections}, this selection is purposefully designed to be conservative, ensuring that detected diffuse \HI emission (i.e clouds in the IGM) is clearly real emission and does not include noise peaks.

However, we found some real \HI emission below these thresholds that should be included in the detection mask. We thus operated on the detection mask using the virtual reality (VR) software \texttt{iDaVIE-v} (Sivitilli et al. in press) from the Institute for Data Intensive Astronomy (IDIA) Visualisation Lab \citep{Marchetti2020, Jarrett2020}. This allowed us to use a ray marching renderer (Comrie et al. in prep) to view and interact with our \HI cube, while making adjustments to the mask within a 3D digital immersive setting. We were able to inspect the mask for any spurious \HI emission that was included or identify real \HI emission that was missed. This was accomplished by importing the detection mask from \SoFiA, overlaying it with the \HI cube in the VR environment, and then adjusting the mask using the motion-tracking hand controllers. As part of this process, we added two sources to the detection mask within the VR environment by marking zones where emission was clearly present. 

The two sources added in VR were originally excluded from the detection mask because they are below the reliability threshold of 0.65 (but above the integrated S/N threshold of 4). These sources are deemed real because they either coincide with emission at other wavelengths (see below) or are part of large, coherent \HI emission. Following these edits to the  detection mask in  VR, we created \HI intensity and velocity maps that are presented in the next section.

\subsection{VST \Ha observation}
\label{sec:Ha_obs}

To generate the \Ha-emission images, we used a combination of \Ha narrow-band images and $r'$ broad-band images both collected using the OmegaCAM attached to the VST at Cerro Paranal, Chile (PID: 0102.B-0780(A)). The OmegaCAM is a 32 CCD wide-field camera with a 1$\deg \times$ 1$\deg$ field of view and a pixel size of 0.21\arcsec. We used the NB 659 \Ha filter with 10\,nm throughput, bought by Janet Drew for the VST Photometric \Ha Survey \citep[VPHAS;][]{Drew2014}. The imaging was done using large $\approx$ 1\,deg pointings and short 150 s and 180 s exposures in $r'$ and \Ha bands, respectively. This strategy allows us to make accurate sky background removal by subtracting averaged background models from the science exposures, and it also reduces the amount of imaging artefacts (such as satellite tracks) in the final mosaics because those are averaged out when the images are stacked. The total exposure times in the $r'$-band and \Ha-band were 8\,250s and 31\,140s, respectively. Similar data reduction and calibration was done for both $r'$-band and \Ha-images. Details of the used reduction steps are given by \citet{Venhola2018}. 

As the \Ha narrow-band images are sensitive both to \Ha emission and flux coming from the continuum, we needed to subtract the continuum flux from the \Ha images before they can be used for \Ha analysis. As the flux in the  $r'$ band is dominated by the continuum, we use scaled $r'$-band images to subtract the continuum from the \Ha. The optimal scaling of the $r'$-band image was selected by visually determining the scaling factor that results in a clean subtraction of the majority of stars and early-type galaxies.

However, there are some caveats in this procedure, which leaves some systematic over- and under-subtraction in the \Ha images. If the seeing conditions or point spread functions (PSFs) differ between the broad- and narrow-band images there will be some residuals in the continuum subtracted image. In addition to these residuals caused by the inner parts of the PSF ($\lesssim$ 5\arcsec), also the extended outer parts \citep[see][]{Venhola2018} and reflection rings of the PSF may leave some features in the images. In the case of bright, extended, and peaked galaxies such as NGC\,1316, these PSF features are also significant. As the positions of the reflection rings are dependent on the position of the source on the focal plane they do not overlap precisely in the narrow- and broad-band images and thus leave some systematic over- and under- subtractions in the images. These kinds of features are apparent in the reduced \Ha emission images.

The over- and under-subtraction artefacts dominate in and around objects with bright stellar emission. Therefore, NGC\,1316 is significantly affected to the extent that the artefacts obscure real \Ha emission. To rectify this, we select a sub-region that includes NGC\,1316, NGC\,1317, and NGC\,1310 and create a model of the background that is ultimately subtracted from the original image. 

The background model was created by masking the visible, real \Ha emission, and replaced with the background local median. The masked image is then filtered with a median filter to eliminate sharp features in the image. Lastly, the (masked, filtered) background model is subtracted from the original image. 

We repeat this process using the residual image to create an improved mask, which is then subtracted from the original image. We use a conservative approach to mask the \Ha emission, as the aim is to remove the dominant artefacts and achieve a uniform background throughout the image. We present a comparison of the images and additional detail in Appendix \ref{sec:appen}.

%--------------------------------------------------------------------

\section{\HI distribution in the group}
\label{sec:results}
\begin{figure*}

    \includegraphics[width = \textwidth]{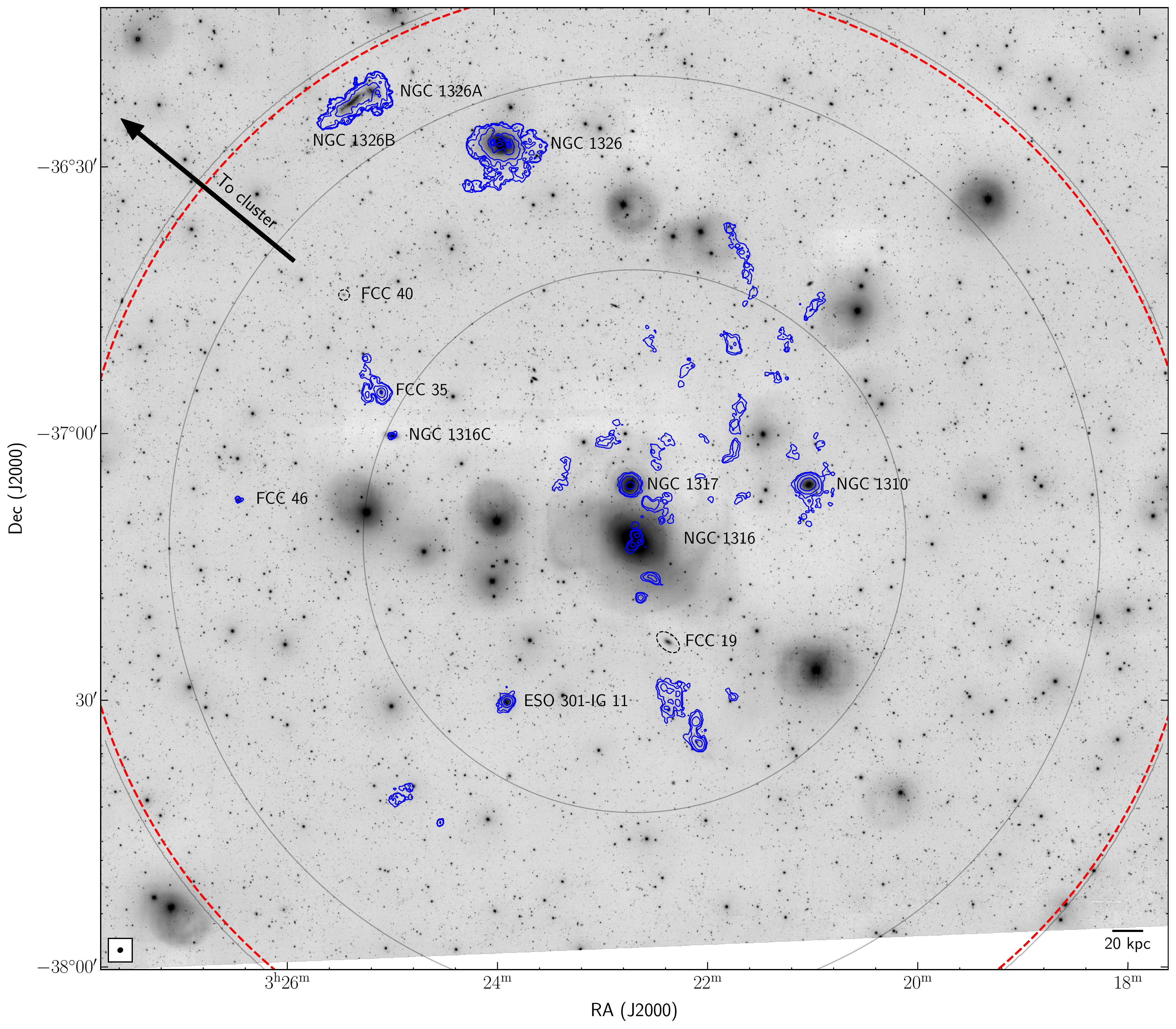}

\caption{Primary beam-corrected constant \HI contours from MeerKAT (blue) overlaid on a FDS \citep{Iodice2016} $gri$ stacked optical image. The lowest contour represents the 3$\sigma$ column density level of N$_{\HI}$ = 1.4 $\times$ 10$^{19}$ atoms cm$^{-2}$ over a 44.1 \kms channel, where the contours increase by a factor of 3$^{n}$ ($n$ = 0, 1, 2, ...). The group galaxies are labelled and the galaxies not detected in \HI are outlined by a dashed black ellipse. The grey circles indicate the sensitivity of the primary beam \citep{Mauch2020} at 50\%, 10\%, and 2\%. The red dashed circle denotes the 1.05 degree (0.38\,Mpc) virial radius of the group as adopted in \cite{Drinkwater2001}, where the restoring beam (33.0\arcsec $\times$ 29.2\arcsec) is shown in the bottom left corner and a scale bar indicating 20\,kpc at the distance of Fornax\,A in the bottom right corner. The direction to the Fornax cluster is shown by the black arrow. In \HI, we detect 10 (out of 12) galaxies, previously known clouds and streams in the IGM and a population of new \HI clouds in the IGM. The previously known IGM \HI structures are labelled in Fig. \ref{fig:mom1} for clarity.}
\label{fig:optical_mom0}
\end{figure*}

In Fig. \ref{fig:optical_mom0}, we present the primary beam-corrected \HI column density map as detected by MeerKAT, overlaid on a $gri$ stacked optical image from the FDS \citep{Iodice2016, Iodice2017, Venhola2018}. Our \HI image (Fig. \ref{fig:optical_mom0}) is sensitive to a column density of N$_{\HI}$ = 1.4 $\times$ 10$^{19}$ atoms cm$^{-2}$ in the most sensitive part (pointing centre), equating to a 3$\sigma$ \HI mass lower detection limit 1.7 $\times$ 10$^{6}$ \Msol for a point source 100\kms wide. 

As a result of the improved sensitivity of our image, in \HI we detect 10 galaxies out of the 13 spectroscopically confirmed galaxies \citep{Maddox2019}, all the previously known clouds and streams, and a new population of clouds and streams in the IGM. Eleven of our \HI detections (10 galaxies and SH2) have corresponding optical redshifts \citep{Maddox2019}. NGC\,1341, FCC\,19, and FCC\,40 are the 3 galaxies with optical redshifts in which we do not detect any \HI. NGC\,1341 is a late-type (SbcII) galaxy with a stellar mass of 5.5 $\times$ 10$^{9}$ \Msol \citep{Raj2020}, in which \HI has beeen previously detected \citep{Courtois2015}. However, NGC\,1341 is outside our \HI image field of view and we do not include it in our sample. FCC\,40 is a low surface brightness dwarf (dE4) elliptical \citep{Iodice2017} and is unlikely to contain massive amounts of \HI. It is also located in a region of the image in which the sensitivity is 75\% worse than the pointing centre, such that we do not detect \HI below 5.6 $\times$ 10$^{19}$ atoms cm$^{-2}$. FCC\,19 is a dS0 with a stellar mass of 3.4 $\times$ 10$^{8}$ \Msol \citep{Iodice2017, Liu2019}. As it is near the pointing centre (70\,kpc in projection from NGC\,1316), we would expect to detect \HI if there were any. However, no \HI is detected in FCC\,19 and we discuss the implications of this in section \ref{sec:cool_gas}. 

\begin{figure*}

    \includegraphics[width = 0.97\textwidth]{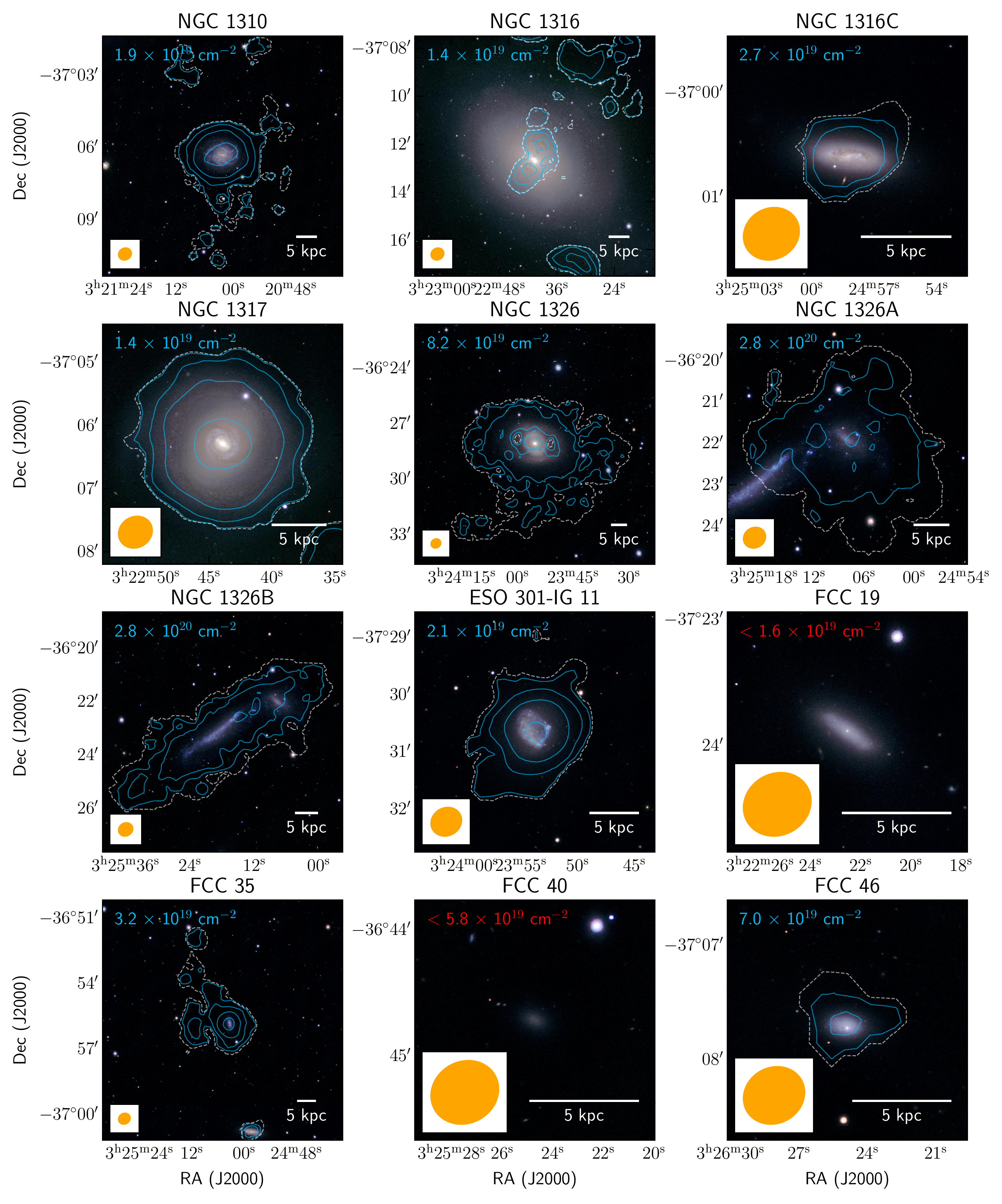}

\caption{Optical three-colour composite of each group galaxy in our sample with overlaid \HI contours. The colour image is comprised of the $g$-, $r$-, and $i$-band filters from the FDS \citep{Iodice2016}; the white dashed contour shows the most sensitive, constant column density of N$_{\HI}$ = 1.4 $\times$ 10$^{19}$ atoms cm$^{-2}$ from Fig. \ref{fig:optical_mom0} and the blue contours start from the local column density sensitivity (i.e. 1.4 $\times$ 10$^{19}$ atoms cm$^{-2}$ scaled by the primary beam response; see top left corner of each cutout) and increase by a factor of 3$^{n}$ with $n$ = 0, 1, 2, ..., at each step. For non-detections, the 3$\sigma$ \HI column density upper limit over a single channel is shown in red in the top left of the cutout. The restoring beam (33.0\arcsec $\times$ 29.2\arcsec) is shown in orange in the bottom left corner and a 5\,kpc scale bar is shown in the bottom right corner.}
\label{fig:cutouts}
\end{figure*}

\begin{figure*}

    \includegraphics[width = \textwidth]{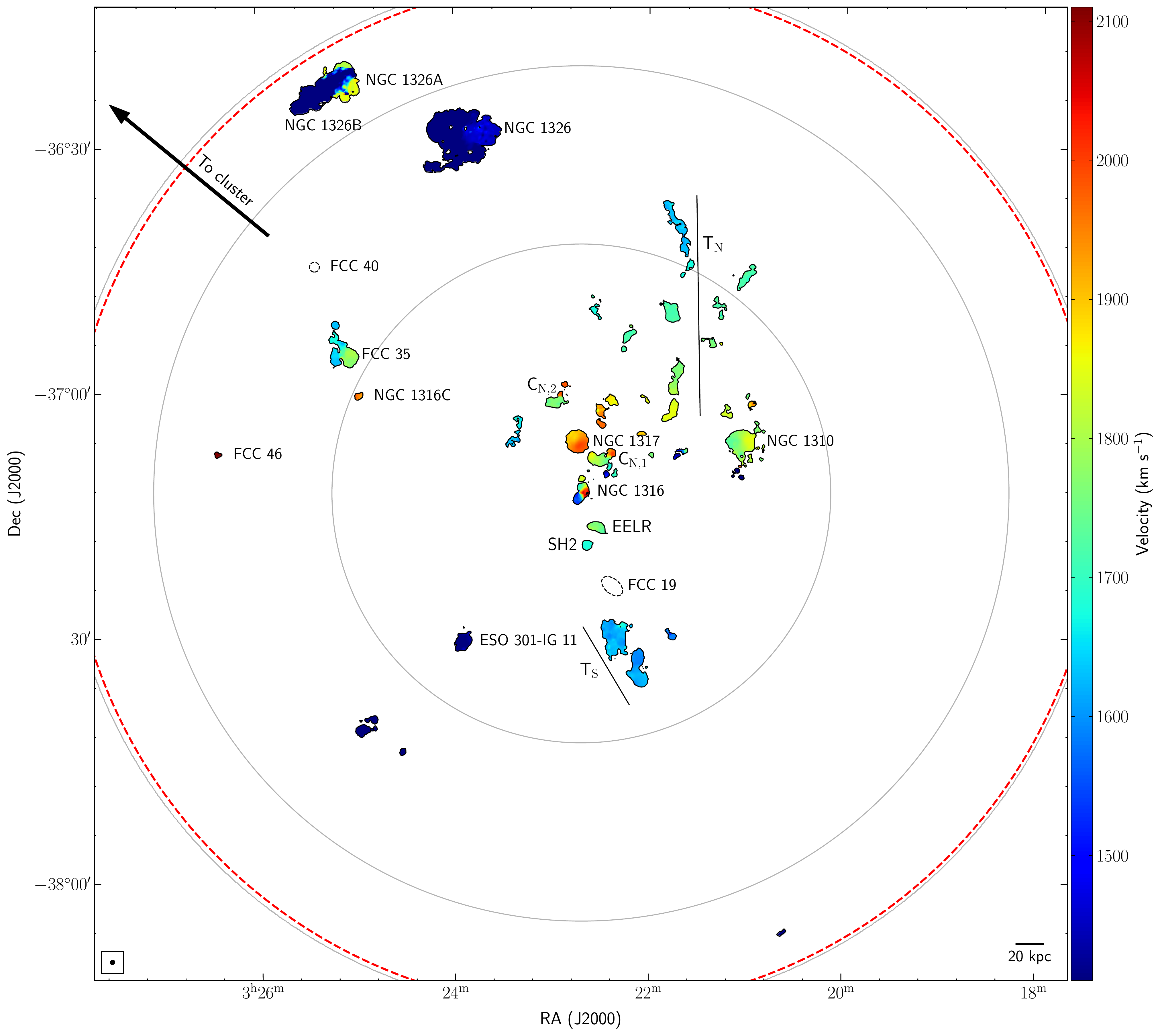}

\caption{\HI velocity field, where the known galaxies and previously detected clouds and tails in the IGM are labelled. As in Fig. \ref{fig:optical_mom0}, the two galaxies not detected in \HI are outlined by black, dashed ellipses and the direction to the Fornax cluster is shown by the black arrow. The velocity colour bar is centred on the systemic velocity of the BGG (NGC\,1316) at 1760\kms. The grey circles indicate the sensitivity of the primary beam \citep{Mauch2020} at 50\%, 10\%, and 2\%. The red dashed circle denotes the 1.05 degree (0.38\,Mpc) virial radius of the group as adopted in \cite{Drinkwater2001}, where the restoring (33.0\arcsec $\times$ 29.2\arcsec) beam and scale bar are shown in the bottom corners. The clouds that make up T$_{\rm N}$ have a new, extended component, effectively doubling the size compared to its original discovery in \citet{Serra2019}.}
\label{fig:mom1}
\end{figure*}

We present the three-colour (constructed using the individual $g$, $r$, and $i$ images) FDS \citep{Iodice2016} optical image cutout for each group galaxy in our sample, which has been overlaid with the \HI contours at their respective column density sensitivity (or upper limit) in Fig. \ref{fig:cutouts}. The integrated \HI flux and mass of the \HI detections and the basic properties of the group galaxies within the \HI image field of view are presented in Table \ref{tab:HI_prop}. The velocity field is presented in Fig. \ref{fig:mom1} and highlights some new large-scale coherent \HI structures, which extend up to $\sim$ 220\,kpc in length. 

\begin{table*}
  \begin{center}
    \caption{Basic properties of the group galaxies and \HI detected sources within the \HI image field of view. The primary beam-corrected integrated \HI flux, mass, and upper limits are included for all sources while the morphological type, stellar mass, and $g$ -- $r$ colour is included for all the galaxies. The \HI mass was calculated using a distance of 20\,Mpc and the statistical uncertainty of the flux was measured and propagated to the \HI mass. The 3$\sigma$ upper limits of the \HI flux and mass are calculated for non-detections using the local RMS and a 100\kms wide integrated flux for a point source. All previously known sources are individually identified and the remaining \HI IGM detections are summed into the remaining clouds category. The galaxy morphologies are classified in \citet{Ferguson1989}, the photometry is used to estimate the stellar mass \citep[with the method of][]{Taylor2011}, and $g$ - $r$ colours are measured in \citet{Raj2020} for the majority of the galaxies and in \citet{Venhola2018} for FCC\,19 and FCC\,40. The photometry, $g$ - $r$ colour, and stellar mass of NGC\,1316 are measured independently in \cite{Iodice2017}.}
    \label{tab:HI_prop}
    \begin{tabular}{c c c c c c}
        \hline
        \hline
        Source & Integrated flux & \HI mass & Morphological type & Stellar mass & $g$ -- $r$ \\
         & (Jy \kms) & (10$^{7}$ \Msol) & & (10$^{9}$ \Msol) & (mag) \\
        \hline        
        NGC\,1310 & 5.13 $\pm$ 0.07 & 48.1 $\pm$ 0.6 & SBcII & 4.7 & 0.6 $\pm$ 0.1 \\
        NGC\,1316 & 0.72 $\pm$ 0.04 & 6.8 $\pm$ 0.4 & SAB0 & 600 & 0.72 $\pm$ 0.01 \\
        NGC\,1316C & 0.18 $\pm$ 0.02 & 1.7 $\pm$ 0.2 & SdIII pec & 1.4 & 0.7 $\pm$ 0.1 \\
        NGC\,1317 & 2.96 $\pm$ 0.02 & 27.8 $\pm$ 0.2 & Sa pec & 17.1 & 0.77 $\pm$ 0.02 \\
        NGC\,1326 & 24.3 $\pm$ 0.5 & 228 $\pm$ 4 & SBa(r) & 29.4 & 0.62 $\pm$ 0.04 \\
        NGC\,1326A & 15.2 $\pm$ 0.8 & 142 $\pm$ 8 &  SBcIII & 1.7 & 0.5 $\pm$ 0.1 \\
        NGC\,1326B & 49 $\pm$ 1 & 455 $\pm$ 9 & SdIII & 1.8 & 0.3 $\pm$ 0.1 \\
        ESO\,301-IG\,11 & 1.52 $\pm$ 0.04 & 14.3 $\pm$ 0.4 & SmIII & 2.9 & 0.57 $\pm$ 0.04 \\
        FCC\,19 & $<$ 0.03 & $<$ 0.17 & dS0 & 0.18 & 0.62 $\pm$ 0.04 \\
        FCC\,35 & 3.51 $\pm$ 0.09 & 33.0 $\pm$ 0.8 & SmIV & 0.17 & 0.2 $\pm$ 0.1 \\
        FCC\,40 & $<$ 0.15 & $<$ 0.72 & dE4 & 0.002 & 0.61 $\pm$ 0.04 \\
        FCC\,46 & 0.13 $\pm$ 0.03 & 1.2 $\pm$ 0.2 & dE4 & 0.58 & 0.46 $\pm$ 0.01 \\
        T$_{\rm N}$ & 2.24 $\pm$ 0.07 & 21.0 $\pm$ 0.7 & - & - & - \\
        T$_{\rm S}$ & 4.86 $\pm$ 0.08 & 45.6 $\pm$ 0.7 & - & - & - \\
        C$_{\rm N,1}$ & 0.75 $\pm$ 0.05 & 7.0 $\pm$ 0.5 & - & - & - \\
        C$_{\rm N,2}$ & 0.35 $\pm$ 0.03 & 3.3 $\pm$ 0.3 & - & - & - \\
        EELR & 0.49 $\pm$ 0.02 & 4.6 $\pm$ 0.2 & - & - & - \\
        SH2 & 0.31 $\pm$ 0.02 & 2.9 $\pm$ 0.2 & - & - & - \\
        Remaining clouds & 3.0 $\pm$ 0.2 & 28 $\pm$ 2 & - & - & - \\

        \hline
    \end{tabular}
  \end{center}
\end{table*}

\subsection{Newly detected \HI}
 Our \HI image is the widest and deepest interferometric image of the Fornax\,A group to date. Naturally, we detect new \HI sources, additional \HI in known sources and resolved \HI in previously unresolved sources. All the sources are presented in Table \ref{tab:HI_prop}, Fig. \ref{fig:optical_mom0}, and \ref{fig:mom1}. As described in Section \ref{sec:group_desc}, several sources in the Fornax\,A group have been previously detected. The new \HI sources detected in this work are as follows:\ resolved \HI tails associated with FCC\,35,  NGC\,1310, and NGC\,1326; an extension of T$_{\rm N}$ in the form of additional, coherent clouds; an additional component to T$_{\rm S}$ in the form of a western cloud; and a population of clouds in the IGM (unlabelled in Fig. \ref{fig:mom1}).

\subsection{\HI in galaxies}
We detect \HI in ten galaxies, where the \HI is well resolved in eight of them (Fig. \ref{fig:cutouts}). Out of those, two galaxies have \HI that is confined to the stellar disc, while the remaining 6 have \HI emission that extend beyond the stellar disc. The two galaxies with unresolved \HI are NGC\,1316C and FCC\,46. 

The two well-resolved galaxies with \HI confined within the stellar discs are NGC\,1316 and NGC\,1317 (Fig. \ref{fig:cutouts}). We detected 6.8 $\times$ 10$^{7}$ \Msol of \HI in the centre of NGC\,1316, 60\% more \HI in the centre than previously detected in \citep{Serra2019}. The \HI has complex kinematics (also seen in the molecular gas and dust) beyond a uniformly rotating disc. The \HI in NGC\,1317 is sharply truncated at the boundary of the stellar disc. Given its stellar mass and morphology, NGC\,1317 is \HI deficient by at least an order of magnitude (discussed in detail in section \ref{sec:cool_gas}).

There are six galaxies in the group that have extended \HI discs. Three galaxies (NGC\,1326A/B and ESO\,301-IG\,11) have slightly extended and mostly symmetric \HI discs, while the other three galaxies (FCC\,35, NGC\,1326, and NGC\,1310) have extended \HI features that are significantly disturbed and asymmetric (Fig. \ref{fig:cutouts}). 

NGC\,1326A and B have extended \HI discs and although they overlap in projection, they are separated by $\sim$ 800 \kms in velocity. There is no \HI connecting these two galaxies along the line of sight down to a column density of 2.8 $\times$ 10$^{20}$ atoms cm$^{-2}$, which is also confirmed through visual inspection in virtual reality. Future, more sensitive data from the MFS \citep{Serra2016} will unambiguously show whether these galaxies are interacting or not.

The collisional ring galaxy ESO\,301-IG\,11 has a slightly extended \HI disc, where the extension is in the south-east direction (away from the group centre). As suggested by its classification, the \HI is likely to have been tidally disturbed in the collision that formed the ring.

In the three galaxies with disturbed or asymmetric \HI discs (detailed below), strong tidal interactions can be reasonably excluded as the cause, as the deep $g$-band FDS images show no stellar emission associated with the extended \HI down to a surface brightness of 30\,mag\,arcsec$^{-2}$. The \HI tails and asymmetries all differ in these galaxies, likely because each galaxy is affected by different processes, such as gentle tidal interactions, ram pressure, and accretion.

The dwarf late-type galaxy FCC\,35 has a long, asymmetric (kinematically irregular) \HI tail pointing away from the group centre. The two closest galaxies (spatially with confirmed redshifts) are NGC\,1316C and FCC\,46, a dwarf late type and dwarf early type. These two galaxies have unresolved \HI and are more \HI deficient than the majority of the group galaxies. Neither a dynamical interaction between these galaxies nor a hydrodynamical mechanism (such as ram pressure) can be ruled out as the cause for the long, \HI tail of FCC\,35.

NGC\,1326 is a barred spiral galaxy with a ring and has clumpy, extended, and asymmetric \HI emission in the south, pointing towards the group centre. The one-sided \HI emission could be indicative of a tidal interaction. However, this could also be an instrumental effect, as the galaxy is located very far from the pointing centre and is subjected to a variable sensitivity response. The southern side (where the \HI tail is) is sensitive down to $\sim$ 6.1 $\times$ 10$^{19}$ atoms cm$^{-2}$, while the northern side has a lower sensitivity of $\sim$ 2.3 $\times$ 10$^{20}$ atoms cm$^{-2}$. As the tails are diffuse ($<$ 1 $\times$ 10$^{20}$ atoms cm$^{-2}$), more sensitive observations are needed to determine if NGC\,1326 has extended \HI emission on the northern side.

Finally, the massive late-type galaxy NGC\,1310 is surrounded by \HI extensions and clouds of different velocities, which is unusual, because it is  a relatively (compared to NGC\,1317) isolated galaxy, with an undisturbed optical spiral morphology and a uniformly rotating \HI disc. Despite the coarse velocity resolution, we can determine from our observations that the majority of the \HI extensions and clouds (except for the extended component of the disc to the south) are not rotating with the disc (Fig. \ref{fig:mom1}) and cover a broad range ($\sim$ 1450 -- 1950\kms) in velocity, suggesting that it may be anomalous \HI gas from an external origin. Future data from the MFS \citep{Serra2016} with better velocity resolution will clarify this point.

\subsection{\HI in the intra-group medium}
\label{sec:res_IGM}
We detect a total of (1.12 $\pm$ 0.02) $\times$ 10$^{9}$ \Msol of \HI in the IGM. All of the previous clouds in \citet{Serra2019} were detected as well as additional \HI in some of these features. We detect new clouds, the majority residing in the north, with some forming large, contiguous 3D structures. 

\begin{figure}

    \includegraphics[width = \columnwidth]{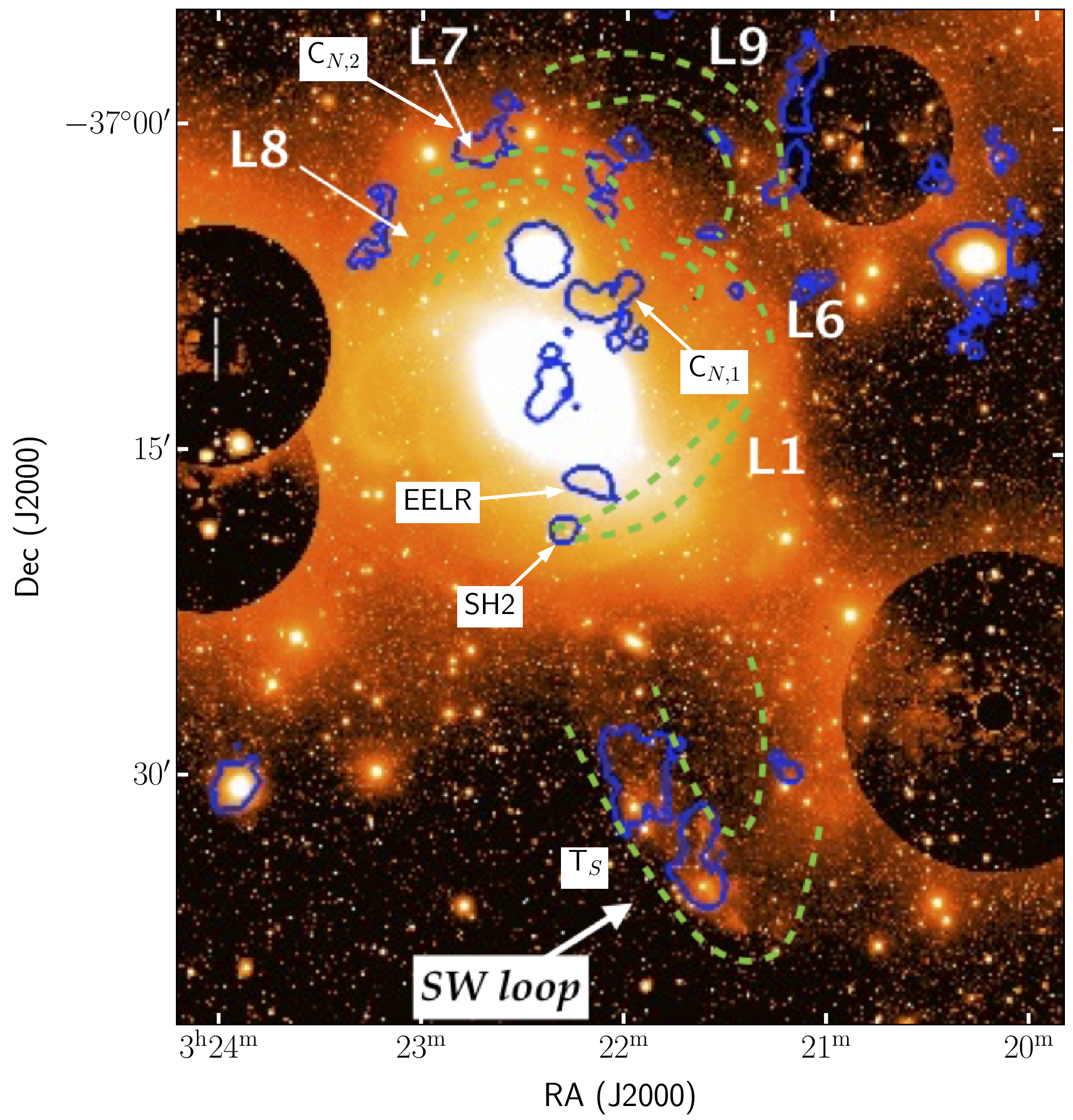}

\caption{Low surface brightness (star removed) image of NGC\,1316 in $g$-band, observed with the VST \citep{Iodice2017}. The known \citep{Schweizer1980, Richtler2014, Iodice2017} stellar loops are labelled and outlined by the dashed green lines. The \HI is shown by the solid blue contours and the previously known \HI clouds are labelled. The clouds that make up T$_{\rm S}$ (including the new western \HI cloud) overlap with the stellar SW loop. There is some overlap with some \HI clouds in the north (e.g. C$_{\rm N, 2}$ and the clouds to the west) and the optical loop L7. Overall, there is no consistent correlation between the stellar loops and the distribution of \HI clouds.}

\label{fig:HI_loops}
\end{figure}

We searched for any association between the new \HI in the IGM and stellar emission. In particular, as more \HI has been detected within the stellar halo of NGC\,1316, we checked for any correlation between the \HI and known stellar loops (Fig. \ref{fig:HI_loops}). Overall, there is very little, clear association between the \HI in the NGC\,1316 halo and its stellar loops. The major exceptions are T$_{\rm S}$ and its newly detected cloud, as they are fully contained within the SW stellar loop. The \HI in SH2 and EELR may potentially correlate with the stellar loop L1 and there are some \HI cloud (e.g. C$_{\rm N, 2}$) in the north that partially overlap with the stellar loop L7. Other than examples above, all the remaining \HI in the IGM shows no association with stellar emission.

We detect an extension in T$_{\rm N}$, effectively doubling its length and mass. The extension smoothly connects in velocity with the previously known emission and now extends up to $\sim$ 220\,kpc from NGC\,1316 (Fig. \ref{fig:mom1}), which is where the \HI originated from \cite{Serra2019}. The clouds that make up T$_{\rm N}$ now contains (2.10 $\pm$ 0.07) $\times$ 10$^{8}$ \Msol of \HI. The north and south tails contain 60\% (6.7 $\pm$ 0.1 $\times$ 10$^{8}$ \Msol) of the total IGM \HI mass. The remaining clouds in the IGM mostly reside to the north of NGC\,1316, with the majority of these existing over a narrow (90\kms) velocity range. It is possible some of these clouds form large coherent \HI structures, although it is not clear compared to the case of T$_{\rm N}$ and T$_{\rm S}$. While T$_{\rm N}$ and T$_{\rm S}$ originate from a single pericentric passage of the NGC\,1316 merger \citep{Serra2019}, the remaining clouds in the IGM are more likely to be the remnants of recently accreted satellites onto NGC\,1316, which is consistent with \citet{Iodice2017}.

The clouds immediately to the north-west of NGC\,1317 may be a remnant of its outer disc. These clouds are within a projected distance of 10\,kpc from NGC\,1317 and the cloud and the galaxy have the same velocity. The \HI-to-stellar mass ratio of the galaxy is low by at least an order of magnitude (see below) and these clouds alone are not enough to explain the \HI deficiency. However, these are the only clouds that show potential evidence that they originated from NGC\,1317.

All the \HI in the IGM located north of the group centre (NGC\,1316) and the clouds to the south-east of ESO\,301-IG\,11 appear to be decoupled from the stars. The \HI in the south (SH2, T$_{\rm S}$) has stellar emission associated with it. Additionally, there are a few \HI clouds near to the group centre that contain multiphase gas. 

\subsection{Multiphase gas in the intra-group medium}
\begin{figure*}

    \includegraphics[width = \textwidth]{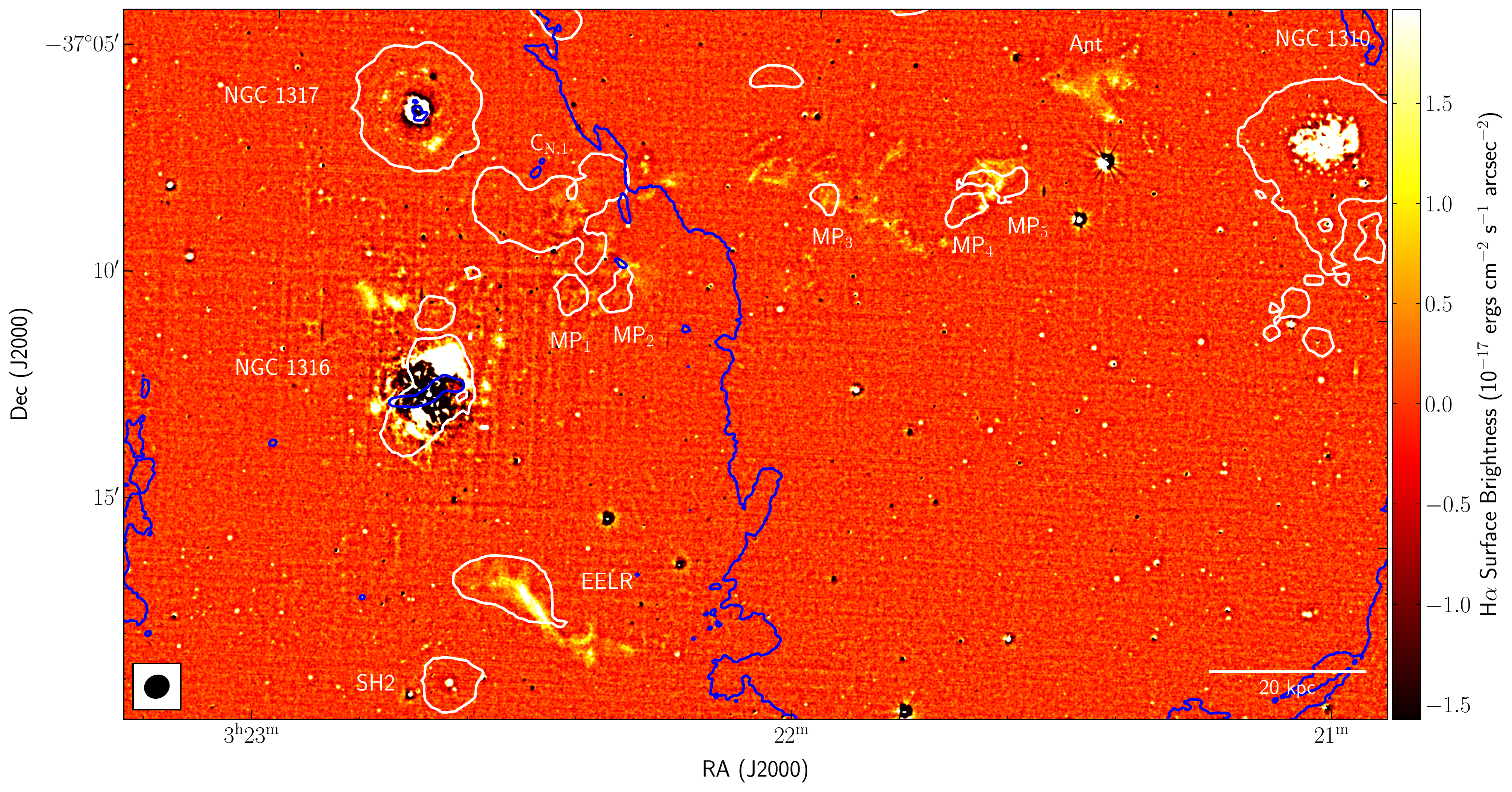}

\caption{OmegaCAM \Ha emission showing the ionised gas in the vicinity of NGC\,1316. The blue contour shows the majority of the western lobe of NGC\,1316 in radio continuum at a (conservative) level of 1.3\,mJy beam$^{-1}$ from \citet{Maccagni2020}. The white contours show the 3$\sigma$ \HI column density of 1.4 $\times$ 10$^{19}$ atoms cm$^{-2}$ (over 44.1\kms) from this work. Known sources (i.e. galaxies and IGM \HI) and multiphase (MP) gas clouds that contain \Ha and \HI as well as the Ant-like feature from \citet{Formalont1989} are labelled. This image reveals long filaments of ionised gas in the IGM.}

\label{fig:Halpha}
\end{figure*}

In Figure \ref{fig:Halpha}, we show the ionized \Ha gas emission detected in the vicinity of NGC\,1316 (i.e. the group centre). \Ha is detected in NGC\,1316, NGC\,1317, and NGC\,1310. However, the most striking features are the \Ha complexes detected in the IGM.

% Can we measure Halpha masses?

There are giant filaments of \Ha in the IGM stretching between galaxies of the group. \HI is directly associated with some of the ionised gas, showing the coexistence of multiphase gas in the IGM. These occur in EELR, C$_{\rm N, 1}$, the cloud directly below C$_{\rm N, 1}$ and in five newly detected clouds containing \HI that we label MP in Fig.\,\ref{fig:Halpha}. Additionally, we detect the ``Ant'' (or ALF; Ant-like feature) first detected as a depolarising feature in \citet{Formalont1989} and later in \Ha by \citet{Bland-Hawthorn1995}. The \Ha emission is thought to provide the intervening turbulent magneto-ionic medium required to depolarize the radio continuum emission \citet{Formalont1989}. There is no optical continuum emission nor any \HI emission currently associated with the Ant.

While there are a number of multiphase gas clouds in the IGM, the brightest case is EELR. It is clear that EELR has a complex multiphase nature, with \HI, \Ha, and dust all previously detected in it \citep{Mackie1998, Horellou2001, Serra2019}. We detect 50\% more \HI than the previous study \citep{Serra2019} and \HI is only present in the region of the bright, more ordered ionised gas morphology. Given that our \HI image is sensitive to a column density of 1.4 $\times$ 10$^{19}$ atoms cm$^{-2}$, it is unlikely that there is any \HI in the less ordered (and likely turbulent) part of EELR. Currently, the origin of EELR is unclear, and we will present a detailed analysis of it and its multiphase gas in future work.

%--------------------------------------------------------------------

\section{Pre-processing in the group}
\label{sec:discussion}
The Fornax\,A group is at a projected distance of $\sim$ 1.3\,Mpc (approximately 2 virial radii) from the Fornax Cluster centre. Redshift independent distances are too uncertain to establish whether the group is falling into the cluster.  However, the intact spiral morphologies of group galaxies imply that the group has not passed the cluster pericentre as spiral morphologies do not typically survive more than one pericentric passage \citep[e.g.][]{Calcaneo-Roldan2000}. At this distance, the intra-cluster medium (ICM) of the Fornax cluster should not have a significant impact on the group galaxies, meaning that quenched galaxies are a result of pre-processing within the group. 

An optical analysis of the radial light profiles of the group galaxies and the intra-group light (IGL) concluded that the Fornax\,A group is in an early stage of assembly \citep{Raj2020}. This is evident from the low level (16\%) of IGL and from the group being dominated by late types with undisturbed morphologies and comparable stellar masses \citep{Raj2020}.

In this work, we detect \HI throughout the Fornax\,A group both in the galaxies and the IGM. While the galaxies range from being \HI rich to extremely \HI deficient, the majority of the galaxies contain a regular amount of \HI for their stellar mass. This is consistent with the group being in an early phase of assembly, as the majority of galaxies would be \HI deficient for a group in the advanced assembly stage. The \HI detections show evidence of pre-processing in the form of (2.8 $\pm$ 0.2) $\times$ 10$^{8}$ \Msol of \HI in the IGM, \HI deficient galaxies, truncated \HI discs, \HI tails, and asymmetries. The diversity of galaxy \HI morphologies suggest that we are observing galaxies at different stages of pre-processing, as we detail below.

\subsection{NGC\,1316 merger}
\label{sec:pre_FA}
The most obvious case of pre-processing in the group is NGC\,1316, the BGG. It is a peculiar early type that is the brightest galaxy in the entire Fornax cluster volume and the result of a 10:1 merger that occurred 1 -- 3\,Gyr ago between a massive early-type galaxy and a gas-rich late-type galaxy \citep{Schweizer1980, Mackie1998, Goudfrooij2001, Iodice2017, Serra2019}. There are large stellar loops and streams, an anomalous amount of dust and molecular gas (2 $\times$ 10$^{7}$ and 6 $\times$ 10$^{8}$ \Msol, respectively) in the centre, as well as \HI in the centre and in the form of long tails \citep{Draine2007, Lanz2010, Galametz2012, Morokuma-Matsui2019, Serra2019}. 

The \HI mass budget for a 10:1 merger to produce the features observed in NGC\,1316 requires the progenitor to contain $\sim$ 2 $\times$ 10$^{9}$ \Msol of \HI \citep{Lanz2010, Serra2019}. Recently, \citet{Serra2019} detected 4.3 $\times$ 10$^{7}$ \Msol of \HI in the centre of NGC\,1316, overlapping with the dust and molecular gas, and a total \HI mass of 7 $\times$ 10$^{8}$ \Msol when including the tails and nearby \HI clouds. While these authors detected an order of magnitude more \HI than previous studies, this is a factor of $\sim$ 3 lower than expected. In this work, we detect a \HI mass in the centre of (6.8 $\pm$ 0.4) $\times$ 10$^{7}$ \Msol and a total \HI mass 0.9 -- 1.2\footnote{The lower limit was determined by only including the same \HI sources as \citet{Serra2019} and the T$_{\rm N}$ extension, while the upper limit includes the remaining \HI clouds in the IGM} $\times$ 10$^{9}$ \Msol associated with NGC\,1316 in the form of streams and clouds. This brings the observed \HI mass budget even closer to the expected value under the 10:1 lenticular + spiral merger hypothesis -- just within a factor 1.7 - 2.2, which is well within the uncertainties.

Since the merger 1 -- 3\,Gyr ago, NGC\,1316 has been accreting small satellites \citep{Iodice2017}. The satellites may have contributed to the build up of \HI, however, we do not observe any \HI correlated with dwarf galaxies within 150\,kpc of NGC\,1316. Any contributed \HI is second order compared to the initial merger, which is supported by the \HI mass of NGC\,1316 being dominated by the tails. Tidal forces from the initial merger ejected 6.6 $\times$ 10$^{8}$ \Msol of \HI into the IGM in the T$_{\rm N}$ and T$_{\rm S}$ tails alone. The remaining \HI in the IGM is likely to be a combination of gas decoupled from stars in the initial merger and gas from more recently accreted satellites. \HI tidal tails that span hundreds of kpc in galaxy groups have been shown to survive in the IGM for the same timescale (1 -- 3\,Gyr) from when this merger took place \citep{Hess2017}.

\subsection{Pre-processing status of the group galaxies}
\label{sec:cool_gas}
In this section, we identify galaxies at different stages of pre-processing according to their \HI morphology and cool gas (\HI and H$_{2}$) ratios. The categories are as follows: i) early, where a galaxy has yet to experience significant pre-processing; ii) ongoing, for galaxies that currently show signatures of pre-processing; and iii) advanced, for galaxies that have already experienced significant pre-processing.

There are a total of 12 galaxies in the sample, which are all the spectroscopically confirmed galaxies within the \HI image field of view. In our sample, 10 galaxies have \HI detections and 2 galaxies (FCC\,19 and FCC\,40) have \HI upper limits (Fig. \ref{fig:cutouts}). There are 7 galaxies that have been observed with ALMA. The 5 galaxies that were not observed are ESO\,301-IG\,11, FCC\,19, FCC\,35, FCC\, 40, and FCC\,46 \citep[][Morokuma-Matsui et al. in prep]{Morokuma-Matsui2019}. We measure the molecular gas mass of the observed galaxies using the standard Milky Way CO-to-H$_{2}$ conversion factor of 4.36 (\Msol K \kms pc$^{-2}$)$^{-1}$ \citep{Bolatto2013} as well as estimated stellar masses (Table \ref{tab:HI_prop}) from \citet{Raj2020} and \cite{Venhola2018}, which are derived from the $g$ and $i$ photometric relation in \citet{Taylor2011}. We remove the helium contribution from our molecular gas masses so that we are measuring the molecular-to-atomic hydrogen gas mass (except in the total gas fraction, shown below) and can directly compare our findings to \citet{Catinella2018}.

\begin{figure*}

    \includegraphics[width = \textwidth]{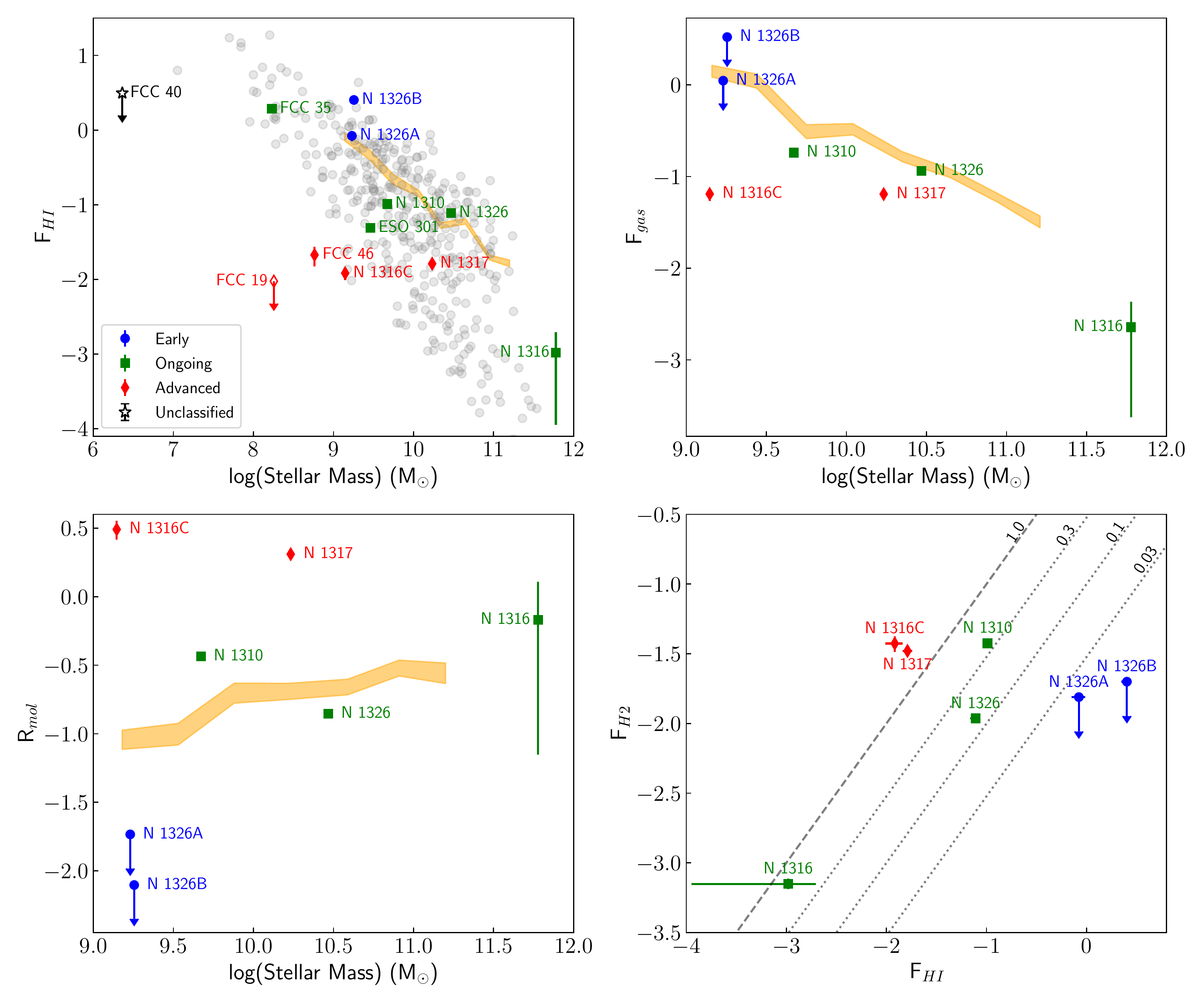}

\caption{Atomic and molecular gas scaling ratios. In all figures, the early, ongoing, and advanced pre-processing categories are shown as blue circles, green squares, and red diamonds, respectively and H$_{2}$ upper limits are depicted by arrows. Solid markers indicate \HI detections and open markers are non-detections. FCC\,40 is not assigned to any pre-processing category and is shown as the open black star. \emph{Top left panel}: The \HI gas fraction compared to galaxies from the HRS \citep{Boselli2010, Boselli2014} and VGS \citep{Kreckel2012} (grey points) that show the typical scatter in \FHI. The orange shaded region indicates the median trend from xGASS \citep{Catinella2018}. \emph{Top right panel}: The total gas fraction of our galaxies compared to the median xGASS-CO trend \citep{Catinella2018} (orange shaded region). \emph{Bottom left panel:} The molecular-to-atomic-gas ratio of our galaxies compared to the median xGASS-CO trend \citep{Catinella2018} (orange shaded region). \emph{Bottom right panel}: The H$_{2}$ gas fraction as a function of \HI gas fraction, showing constant ratios of 100\%, 30\%, 10\%, and 3\%. Overall, the galaxies in the early category are \HI rich, the galaxies in the ongoing category typically follow the xGASS and xGASS-CO median scaling relations \citep{Catinella2018}, while galaxies in the advanced category have no \HI or are \HI-deficient with irregularly high H$_{2}$-to-\HI ratios.}

\label{fig:scaling_rels}
\end{figure*}

We present the \HI and H$_{2}$ scaling ratios in Fig.\,\ref{fig:scaling_rels}. We measure the \HI gas fraction \FHI $\equiv$ log(M$_{\rm HI}$/\Mstar), the total gas fraction Fgas $\equiv$ log(1.3(M$_{\rm HI}$ + M$_{\rm H2}$)/\Mstar) where the 1.3 accounts for the helium contribution, the molecular-to-atomic gas mass ratio R$_{mol}$ $\equiv$ log(M$_{\rm H2}$/M$_{\rm HI}$), and the H$_{2}$ gas fraction \FHmol $\equiv$ log(M$_{\rm H2}$/\Mstar). We compare the \HI fraction of our galaxies to those in the Herschel Reference Survey \citep[HRS;][]{Boselli2010, Boselli2014} and the Void Galaxy Survey \citep[VGS;][]{Kreckel2012}, which span a comparable stellar mass range of our galaxies. We also compare \FHI to the median trend of the extended GALEX Arecibo SDSS Survey \citep[xGASS;][]{Catinella2018}. Furthermore, we compare our molecular gas scaling relations to the median trends of xGASS-CO (Fig.\,\ref{fig:scaling_rels}), which are xGASS galaxies with CO detections \citep{Catinella2018}. The xGASS and xGASS-CO trends provide a good reference for the \HI and H$_{2}$ scaling relations in the local Universe as the median \FHI trend was derived from 1179 galaxies selected with 10$^{9}$ $<$ \Mstar (\Msol) $<$ 10$^{11.5}$ and 0.01 $<$ z $<$ 0.05, and the H$_{2}$ mass and scaling relations derived using a subset 477 galaxies from the parent sample that have CO detections. 

% Perhaps need to say that they used a vairable CO-to-H2 conversion, we used a constant, however their mean value is almost the same as what we use. 

The two galaxies that show no signatures (i.e. in the early phase) of pre-processing are NGC\,1326A and NGC\,1326B. They are \HI rich galaxies with typical extended \HI discs and a low molecular gas content. Both galaxies were observed with ALMA \citep[][Morokuma-Matsui et al. in prep]{Morokuma-Matsui2019}, although no CO was detected, placing upper limits on the H$_{2}$ mass. They have the highest \HI fraction and lowest H$_{2}$-to-\HI ratios given their stellar mass (Fig. \ref{fig:scaling_rels}). The galaxies are just within the virial radius of the group, making them furthest from the group centre in projected distance. This increases the likelihood that the galaxies have not undergone pre-processing yet.

The galaxies that show current signatures of pre-processing (i.e. the ongoing category) are FCC\,35, ESO\,301-IG\,11, NGC\,1310, NGC\,1316, and NGC\,1326. In general, these galaxies have \HI tails or asymmetric extended \HI emission, typical \HI and H$_{2}$ ratios (for the galaxies with H$_{2}$ observations) that follow the median xGASS trends in Fig. \ref{fig:scaling_rels}. The exception to this is NGC\,1316. As this galaxy is the BGG, it has a unique formation and evolution history (discussed in section \ref{sec:pre_FA}) that displays both an ongoing (e.g. tidal tails) and advanced state (giant elliptical with a lack of \HI contained in the stellar body) of pre-processing. In this work, we include this galaxy in the ongoing category, although the \HI mass range calculated in section \ref{sec:pre_FA} reflects that it could also be part of the advanced category.

FCC\,35 is the bluest galaxy (Fig. \ref{fig:cutouts} and Table \ref{tab:HI_prop}) in the group \citep{Raj2020} and has extremely strong and narrow optical emission lines that classify it as either a blue compact dwarf or an active star-burst H$_{\rm II}$ galaxy \citep{Putman1998}. Previous studies \citep[i.e.][]{Putman1998, Schroder2001} detected a \HI cloud associated with FCC\,35 and suggested it may be a result of a tidal interaction with the nearest (projected separation of 50\,kpc) neighbour NGC\,1316C. This is a plausible scenario as FCC\,35 has an up-bending (Type-III) break in the stellar radial profile, and a bluer outer stellar disc \citep{Raj2020}, which could be tidally induced star formation. However, the star formation could also be compression/shock induced \citep{Raj2020}. We detect the \HI cloud of FCC\,35 as part of a long tail pointing away from the group centre, making it the most likely galaxy to show evidence of ram pressure stripping. The lower IGM (compared to the ICM) density means that ram pressure stripping is less prevalent in groups. Despite the observational challenges, a few cases have been reported \citep[e.g.][]{Westmeier2011, Rasmussen2012a, Vulcani2018, Elagali2019} and ram pressure is thought to play an important role in the pre-processing of galaxies in groups. FCC\,35 is not \HI deficient (Fig. \ref{fig:scaling_rels}), implying that the gas has recently been displaced, similar to other galaxies showing early signs of gas removal \citep[e.g.][]{Ramatsoku2020, Moretti2020}. 

ESO\,301-IG\,11 is a collisional ring galaxy with a \HI gas fraction below the median trend, although it is not the most \HI deficient galaxy for its stellar mass. There is clear evidence of a tidal interaction in the form of irregular optical morphology, an up-bending (Type-III) break in the stellar radial profile and a slightly extended and asymmetric \HI disc. The galaxy is blue in colour, although the outer stellar disc is redder than the inner disc \citep{Raj2020}, implying that the tidal interaction may have restarted star formation in the centre.  

The asymmetric \HI tail of NGC\,1326 is diffuse ($<$ 1 $\times$ 10$^{20}$ atoms cm$^{-2}$) and only detected on one side of the galaxy. The sensitivity of the opposing side prevents us from detecting \HI that diffuse, and we are therefore unable to distinguish whether the extended \HI is part of a regular extended \HI disc or a signature of pre-processing. With the current \HI content, it follows the same \HI and H$_{2}$ trends as the other galaxies in the ongoing category.

The optical morphology and gas scaling relations of NGC\,1310 suggest that it is not being pre-processed. The stellar spiral structure is completely intact (Fig. \ref{fig:cutouts}), ruling out strong tidal interactions and the \HI gas fraction and molecular-to-atomic gas ratios are close to the median trends. However, the \HI morphology appears complex and incoherent, with many asymmetric extensions and nearby clouds at different velocities. It is clear that the anomalous \HI clouds and extensions are not rotating with the main \HI disc (Fig. \ref{fig:mom1}), suggesting external origins. The \HI extension in the north-west may be emission from a dwarf satellite galaxy, although a spectroscopic redshift would be required to confirm this. Given the presence of the \Ha filaments in the vicinity of NGC\,1310, the remaining clouds may be a result of hot gas, cooling in the IGM (and hot halo of NGC\,1316) and being captured or accreted onto this galaxy.
% Also has a redder outer-disk. 

Finally, the galaxies that are in the advanced stage of pre-processing are NGC\,1316C, NGC\,1317, FCC\,19, and FCC\,46. There is no \HI detected in FCC\,19, and the other three galaxies have truncated \HI discs and are \HI deficient as their \FHI is more than 3$\sigma$ from the xGASS median trend (Fig. \ref{fig:scaling_rels}). 

NGC\,1316C and NGC\,1317 have a low \HI mass fraction and regular H$_{2}$ mass fraction. The total gas fraction of these galaxies is low and is driven by the lack of \HI. Hence, they have significantly more H$_{2}$ than \HI and a molecular-to-atomic fraction an order of magnitude higher (the highest in our sample) than the median trend (Fig. \ref{fig:scaling_rels}). Both these galaxies have no break (Type-I) in their stellar radial profile \citep{Raj2020}, showing no sign of disruption to their stellar body and their \HI confined to the stellar disc, implying that the outer \HI disc has been removed. Ram pressure or gentle tidal interactions are likely to be responsible for removing the outer \HI disc of these galaxies. The less dense (compared to the ICM) IGM combined with the group potential allows galaxies to hold on to their gas more effectively than in clusters \citep{Ruchika2020}. The retained atomic gas within the stellar body can then be converted into molecular gas. This scenario is consistent with the findings of the GAs Stripping Phenomena in galaxies with MUSE \citep[GASP;][]{Moretti2020} project, where pre-processed galaxies in groups (and clusters) have their outer \HI removed (via ram pressure) and the remaining \HI is efficiently converted into H$_{2}$. These galaxies in the advanced stage of pre-processing with truncated \HI discs and regular amounts of H$_{2}$ are similar to some galaxies in the Virgo \citep{Cortese2010} and Fornax cluster \citep{Loni2021}. This suggests that late-type galaxies that have been sufficiently processed lose their outer \HI disc and end up with more H$_{2}$ than \HI.

Despite the similarities between NGC\,1316C and NGC\,1317, these galaxies have likely been pre-processed on different timescales. The stellar mass of NGC\,1316C is more than an order of magnitude lower than that of NGC\,1317 and according to \citet{Raj2020}, NGC\,1316C only recently ($<$ 1\,Gyr) became a group member while NGC\,1317 may have been a group member for up to 8\,Gyr. There is no star formation beyond the very inner ($<$ 0.5\arcmin) disc of NGC\,1317 \citep{Raj2020} and even though there is only a projected separation of $\sim$ 50\,kpc between NGC\,1316 and NGC\,1317, a strong tidal interaction can be reasonably excluded due to the intact spiral structure of NGC\,1317 \citep{Richtler2014, Iodice2017}. The outer \HI disc has been removed and possibly lost to the IGM (i.e. potentially identified as the adjacent clouds at the same velocity) as a result of gentle tidal or hydrodynamical interactions. Alternatively, the outer disc may have been converted to other gaseous phases on short timescales ($<$ 1 Gyr). While we are unable to identify the exact mechanisms that are responsible for the truncated \HI disc of NGC\,1317, it is evident that the galaxy has not had access to cold gas over long timescales. 

Out of all the galaxies with \HI, FCC\,46 is the most \HI deficient given its stellar mass. It is a dwarf elliptical with a recent star formation event and \HI was first detected as a polar ring orbiting around the optical minor axis by \cite{deRijcke2013}. As the \HI is kinematically decoupled from stellar body, the gas was likely accreted from an external source \citep{deRijcke2013}. Our measured \HI mass (Table \ref{tab:HI_prop}) is consistent with that from \citet{deRijcke2013}, although, as a result of our sensitivity at that position, we do not detect the diffuse \HI component that shows the minor axis rotation. A minor merger event (e.g. with a dwarf late type) is consistent with the morphology and $\sim$ 10$^{7}$ \Msol of \HI found in FCC\,46.

FCC\,19 is a dwarf lenticular galaxy (Fig. \ref{fig:cutouts}) with a stellar mass of 3.4 $\times$ 10$^{8}$ \Msol \citep{Liu2019}. It has a $g$ -- $r$ colour of 0.58 \citep{Iodice2017}, which is similar to the colour of NGC\,1310, NGC\,1326, NGC\,1326A, and ESO\,301-IG\,11 (Table \ref{tab:HI_prop}), which have regular \HI fractions and are likely forming stars. However, no \HI is detected in FCC\,19 and we measure a 3$\sigma$ \FHI upper limit of -2.3 (Fig. \ref{fig:scaling_rels}) assuming a 100\kms line width. FCC\,19 is situated in the most sensitive part of the image, meaning that the galaxy truly does not contain \HI. Being so close (70\,kpc in projection) to NGC\,1316, the tidal field and hot halo of NGC\,1316 are likely to have played significant roles in removing the \HI from FCC\,19. The \HI has likely been stripped from the galaxy and lost to the IGM. The stripped \HI may also be potentially heated and prevented from cooling.

Lastly, we refrain from assigning a category to FCC\,40 because we are unable to ascertain whether the galaxy properties are a result of secular evolution or have been influenced by pre-processing. This galaxy is a low surface brightness (Fig. \ref{fig:cutouts}), low-mass (M$_{\star}$ = 2.3 $\times$ 10$^{6}$ \Msol) blue dwarf elliptical (Table \ref{tab:HI_prop}) with no \HI detected. We place an upper limit on the \HI mass (and \HI fraction), although it is currently unknown if galaxies of this mass, colour, and morphology are expected to contain \HI.

\begin{figure}

    \includegraphics[width = \columnwidth]{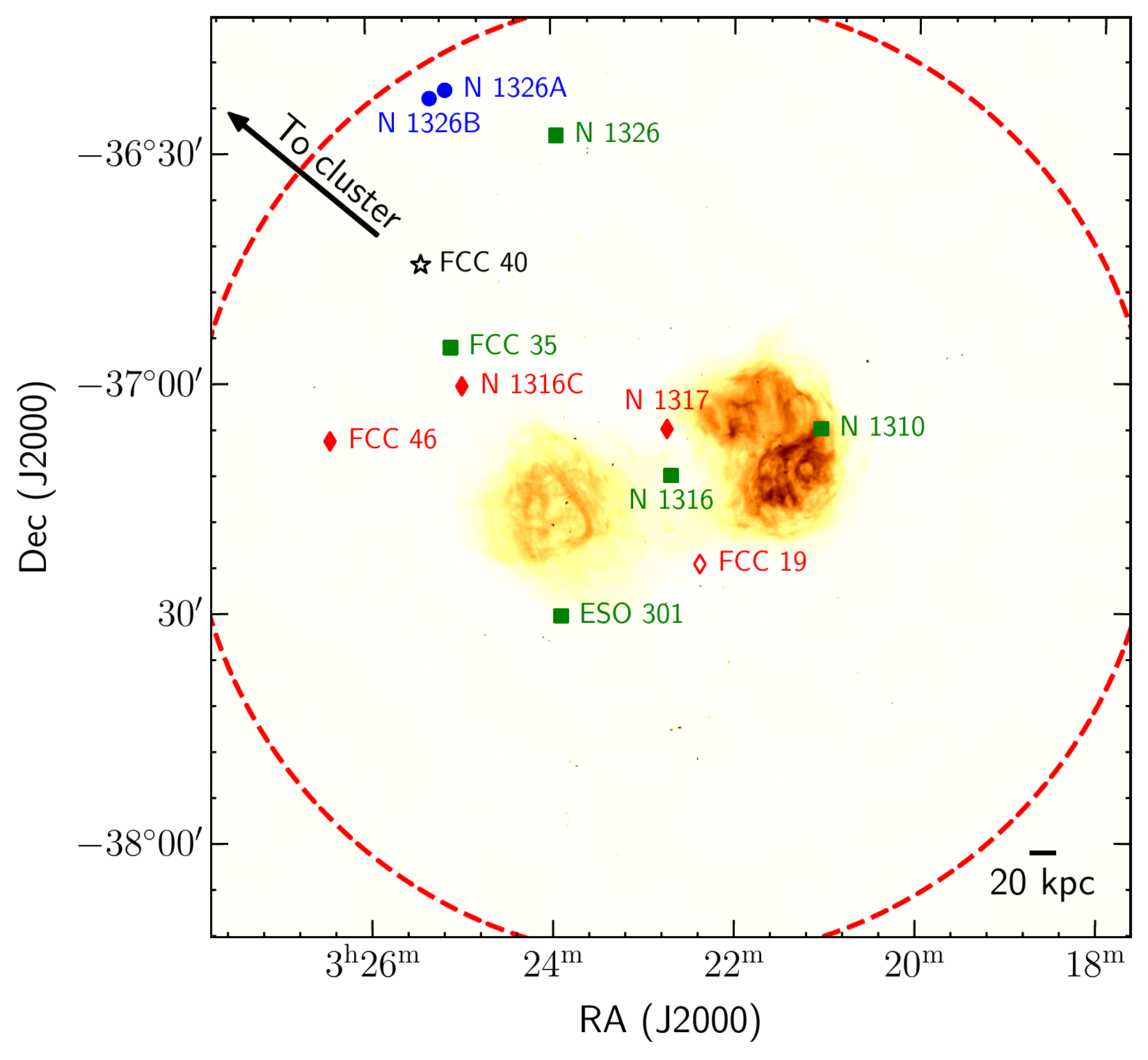}

\caption{Pre-processing map of the Fornax\,A group. The background image shows the 1.44\,GHz MeerKAT radio continuum emission \citep{Maccagni2020} and the position of each group galaxy are overlaid with the same markers as Fig. \ref{fig:scaling_rels}. The filled markers represent \HI detections, the open markers indicate \HI non-detections, where the early, ongoing, advanced, and unclassified pre-processing categories are shown as blue circles, green squares, red diamonds, and black stars, respectively. The red dashed circle denotes the 1.05 degree (0.38\,Mpc) virial radius of the group as adopted in \cite{Drinkwater2001}. A 20\,kpc scale bar is shown in the bottom right corner and the direction to the Fornax cluster is shown by the black arrow. There is no consistent trend between projected position and pre-processing status, although the majority of group galaxies show evidence of pre-processing. The extent of the NGC\,1316 AGN lobes show that it may be playing a role in the pre-processing of neighbouring galaxies and the magnetic field could help the containment of multiphase gas.}

\label{fig:group_map}
\end{figure}

We show the spatial distribution of each group galaxy and their pre-processing status in Fig. \ref{fig:group_map}. The distribution shows a variety of pre-processing stages mixed throughout the group, with no clear radial dependence. The majority of on-going and advanced pre-processing are $<$ 0.5 of the group virial radius, although there are galaxies (i.e. FCC\,46 and NGC\,1326) that have the same pre-processing status and are located closer to edge of the group, $>$ 0.5 of the group virial radius. At a distance of $\sim$ 2 (cluster) virial radii from the Fornax cluster, the Fornax group is located at the distance where  pre-processing is thought to be the most efficient \citep{Lewis2002, Gomez2003, Verdugo2008, Mahajan2012, Haines2015}. In general, it is not clear whether the pre-processing at this infall distance is driven by the group interacting with the cluster, or by local (e.g. tidal and hydrodynamical) interactions within the group. In this instance, it appears that pre-processing is driven by local interactions within the Fornax\,A group for the following reasons: i) The massive, central galaxy is at least one order of magnitude more massive (Table \ref{tab:HI_prop}) than the satellite galaxies. ii) This central galaxy underwent a merger 1 -- 3\,Gyr ago (discussed in Section \ref{sec:pre_FA}). iii) The majority of galaxies close to the group centre ($<$ 0.5 of the group virial radius) show evidence of pre-processing, while the two galaxies (NGC\,1326 A/B) closest to the Fornax cluster (and furthest from the group centre) show no evidence of pre-processing. In addition to these points, there are four galaxies (NGC\,1310, NGC\,1317, ESO\,301-IG\,11, and FCC\,19) that spatially overlap (in projection) with the radio lobes of NGC\,1316 (Fig. \ref{fig:group_map}) and therefore may be influenced by the AGN \citep[e.g][]{Johnson2015}.

\subsection{Gas in the IGM}
The \HI tails and clouds in the IGM are a direct result of galaxies having their \HI removed through hydrodynamical and tidal interactions over the past few Giga-years. As described in sections \ref{sec:res_IGM} and \ref{sec:pre_FA}, the majority (if not all) of the \HI in the IGM is due to the Fornax\,A merger and the recent accretion of satellites. 

The amount (1.12 $\pm$ 0.2 $\times$ 10$^{9}$ \Msol) of detected \HI in the IGM is not enough to account for all of the missing \HI in the \HI deficient group galaxies. However, the outer parts of the image are subject to a large primary beam attenuation and some of the IGM \HI may be hiding in the noise. We estimate the amount of \HI potentially missed by assuming that we detect all \HI in the IGM in the inner 0.1\,deg$^{2}$ (primary beam response $>$ 90\%) and that the IGM \HI in this area is representative of the IGM throughout the entire group both in terms of amount of \HI per unit area and \HI column density distribution. Under this assumption, the primary beam attenuation reduces the detected \HI by a factor of $\sim$ 2.3, implying that we may be missing up to $\sim$ 1.5 $\times$ 10$^{9}$ \Msol of \HI in the IGM. All the (including the missed) \HI in IGM is still not enough to explain all the \HI deficient galaxies in the group and clearly gas exists in other phases (i.e. H$_{2}$ and \Ha). Some of the \HI in the galaxies has been converted into H$_{2}$, which explains why the more advanced pre-processed galaxies that have \HI, display high molecular-to-atomic gas ratios, and there is \Ha in galaxies and the IGM. 

Currently, the origin of giant ionised gas filaments in the IGM is not well understood. However, they are typically observed in high-mass groups or low-mass clusters (e.g. halo masses $>$ 10$^{13.5}$ \Msol), for example the Virgo cluster \citep{Kenney2008, Boselli2018a, Boselli2018b, Longobardi2020} and the Blue Infalling Group \citep{Cortese2006, Fossati2019}; see \citet{Yagi2017} for a list of clusters that contain long ionised gas filaments. A likely scenario is that cool gas is stripped from an in-falling galaxy, and subsequently ionised, possibly from ionising photons originating from star-forming regions \citep{Poggianti2018, Fossati2019} or through non-photo-ionisation mechanisms such as shocks, heat conduction, and magneto-hydrodynamic waves \citep{Boselli2016}. We use the relation in \citet{Barger2013} to estimate the total \Ha mass in the IGM (i.e. EELR, SH2, and the filaments) from our \Ha photometry (Fig. \ref{fig:Halpha}). Assuming a typical \Ha temperature of 10$^{4}$\,K and electron density of 1 cm$^{-3}$, we estimate the total \Ha mass in the IGM to be $\sim$ 2.6 $\times$ 10$^{6}$ \Msol, which does not significantly contribute to the total gas budget in the IGM.

Simulations show that $\sim$ 10$^{4}$\,K (i.e. relatively cool) gas clouds can survive in hot haloes (such as NGC\,1316) for cosmological timescales \citep{Nelson2020}. The clouds originate from satellite mergers, and are not in thermal equilibrium, but rather magnetically dominated. Cooling is triggered by the thermal instability and the cool gas is surrounded by an interface of intermediate temperature gas \citep{Nelson2020}. These ingredients can explain how multiphase gas clouds are present in the hot halo of NGC\,1316 (Fig. \ref{fig:Halpha}), such that the \Ha filaments are a result of satellite accretion and the \HI has rapidly cooled from these structures, with the ability to survive in the IGM for cosmological timescales. 

Recently, \citet{Muller2020} suggest that magnetic fields of the order of 2 -- 4\,$\mu$G can shield \Ha and \HI in the ICM / IGM such that the gas clouds do not dissipate. As the \Ha filaments and multiphase gas clouds are within the radio lobes (in projection) of NGC\,1316, the magnetic field of the lobes \citep[measured to be $\sim$ 3\,$\mu$G by][]{McKinley2015, Anderson2018, Maccagni2020} may be providing additional stability for the \Ha and \HI to survive. Indeed, the Ant detected by \citet{Formalont1989} and \citet{Bland-Hawthorn1995} is a small portion of the giant \Ha filaments in the IGM. Even though there is currently no \HI associated with the Ant, other sections of the \Ha filaments show that neutral and ionised gas can coexist in some regions of the IGM, possibly transform into one another, and accrete onto group galaxies (e.g. NGC\,1310).

\section{Conclusions}
\label{sec:conclusions}
We present results from MeerKAT \HI commissioning observations of the Fornax\,A group. Our observations are reduced with the \texttt{CARACal} pipeline and our \HI image is sensitive to a column density of 1.4 $\times$ 10$^{19}$ atoms cm$^{-2}$ in the field centre. Out of 13 spectroscopically confirmed group members, we detect \HI in 10 and report an \HI mass upper limit for 2 (the remaining galaxy is outside the field of view of our observation). We also detect \HI in the IGM, in the form of clouds, some distributed along coherent structures up to 220\,kpc in length. The \HI in the IGM is the result of a major merger occurring in the massive, central galaxy NGC\,1316, 1 -- 3\,Gyr ago, combined with \HI being removed from satellite galaxies as they are pre-processed.

We find that 9 out of the 12 galaxies show some evidence of pre-processing in the form of \HI deficient galaxies, truncated \HI discs, \HI tails, and asymmetries. Using the \HI morphology and the molecular-to-atomic gas ratios of the galaxy, we classify whether each galaxy is in the early, ongoing, or advanced stage of pre-processing. 

Finally, we show that there are giant \Ha filaments in the IGM, within the hot halo of NGC\,1316. The filaments are likely a result of molecular gas being removed from a satellite galaxy and then ionised. We observe a number of \HI clouds associated with the ionised \Ha filament, indicating the presence of multiphase gas. Simulations show that hot gas can condense into cool gas within hot haloes and survive for long periods of time on a cosmological timescale, which is consistent with the cool gas clouds we detect within the hot halo of NGC\,1316. The multiphase gas is supported by magnetic pressure, implying that the magnetic field in the lobes of the NGC\,1316 AGN might be playing an important role in maintaining these multiphase gas clouds. The cycle of AGN activity and cooling gas in the IGM could ultimately result in the cool gas clouds falling back onto the central galaxy. We summarise our main findings as follows:

   \begin{enumerate}
   
      \item We present new, resolved \HI in FCC\,35, NGC\,1310, and NGC\,1326.
      
      \item There  is a total of(1.12 $\pm$ 0.02) $\times$ 10$^{9}$ \Msol of \HI in the IGM, which is dominated by  T$_{\rm N}$ and T$_{\rm S}$ (combined \HI mass of 6.6 $\times$ 10$^{8}$ \Msol). We detect additional components in both tails, an extension in  T$_{\rm N}$, effectively doubling its length, and a cloud in T$_{\rm S}$ that shows coherence with the stellar south-west loop.
      
      \item The \HI in the IGM is decoupled from the stars, other than in T$_{\rm S}$ and SH2. 
      
      \item We measure 0.9 -- 1.2 $\times$ 10$^{9}$ \Msol of \HI associated with NGC\,1316, bringing the observed \HI mass budget within a factor of $\sim$ 2 of the expected value for a 10:1 lenticular + spiral merger occurring $\sim$ 2\,Gyr ago.
      
      \item Out of the 12 group galaxies in our sample, 2 (NGC\,1326A and NGC\,1326B) are in the early phase of pre-processing, 5 (FCC\,35, ESO\,301-IG\,11, NGC\,1310, NGC\,1316, and NGC\,1326) are in the ongoing phase of pre-processing, 4 (NGC\,1316C, NGC\,1317 FCC\,19, and FCC\,46) are in the advanced stage of pre-processing, and 1 (FCC\,40) remains unclassified. 
      
      \item Galaxies that are yet to be pre-processed have a typical extended \HI disk, high \HI content, and  molecular-to-atomic gas ratios at least an order of magnitude below the median trend for their stellar mass. Galaxies that are currently being pre-processed typically display \HI tails or asymmetric extended disks, while containing regular amounts of \HI and H$_{2}$. Galaxies in the advanced stage of pre-processing have no \HI or have lost their outer \HI and are efficiently converting their remaining \HI to H$_{2}$.

      \item We detect the Ant first observed by \citet{Formalont1989} as a depolarising feature and later in \Ha by \citet{Bland-Hawthorn1995}, which turns out to be a small part of long, ionised \Ha filaments in the IGM. Localised cooling (potentially assisted by the magnetic field in the lobes of the NGC\,1316 AGN) can occur in the \Ha filaments to condense and form \HI.
      
   \end{enumerate}

In this work, our deep MeerKAT \HI image shows many examples of pre-processing in the Fornax\,A group, such as galaxies with a variety of atypical morphologies and massive amounts of \HI in the IGM. The improved sensitivity and resolution of the MFS \citep{Serra2016} will likely reveal more \HI throughout the group and provide kinematic information for the \HI in galaxies and the IGM. 

\begin{acknowledgements}

The MeerKAT telescope is operated by the South African Radio Astronomy Observatory, which is a facility of the National Research Foundation, an agency of the Department of Science and Innovation. We are grateful to the full MeerKAT team at SARAO for their work on building and commissioning MeerKAT. This paper makes use of the following ALMA data: ADS/JAO.ALMA\#2017.1.00129.S. ALMA is a partnership of ESO (representing its member states), NSF (USA), and NINS (Japan), together with NRC (Canada), MOST and ASIAA (Taiwan), and KASI (Republic of Korea), in cooperation with the Republic of Chile. The Joint ALMA Observatory is operated by ESO, AUI/NRAO, and NAOJ.

This work also made use of the Inter-University Institute for Data Intensive Astronomy (IDIA) visualisation lab (\url{https://vislab.idia.ac.za}). IDIA is a partnership of the University of Cape Town, the University of Pretoria and the University of Western Cape.

This project has received funding from the European Research Council (ERC) under the European Union’s Horizon 2020 research and innovation programme (grant agreement no. 679627; project name FORNAX). The research of OS is supported by the South African Research Chairs Initiative of the Department of Science and Innovation and the National Research Foundation. KT acknowledges support from IDIA. The work of KMM is supported by JSPS KAKENHI Grant Number of 19J40004. RFP acknowledges financial support from the European Union’s Horizon 2020 research and innovation program under the Marie Skłodowska-Curie grant agreement No. 721463 to the SUNDIAL ITN network. AV acknowledges the funding from the Emil Aaltonen foundation. PK is partially supported by the BMBF project 05A17PC2 for D-MeerKAT. AS acknowledges funding from the National Research Foundation under the Research Career Advancement and South African Research Chair Initiative programs (SARChI), respectively. FV acknowledges financial support from the Italian Ministry of Foreign Affairs and International Cooperation (MAECI Grant Number ZA18GR02) and the South African NRF (Grant Number 113121) as part of the ISARP RAIOSKY2020 Joint Research Scheme. 

\end{acknowledgements}


\bibliographystyle{aa} % style aa.bst
\bibliography{References} % your references Yourfile.bib

\appendix
\section{\Ha image comparison}
\label{sec:appen}

\begin{figure*}

    \includegraphics[width = \textwidth]{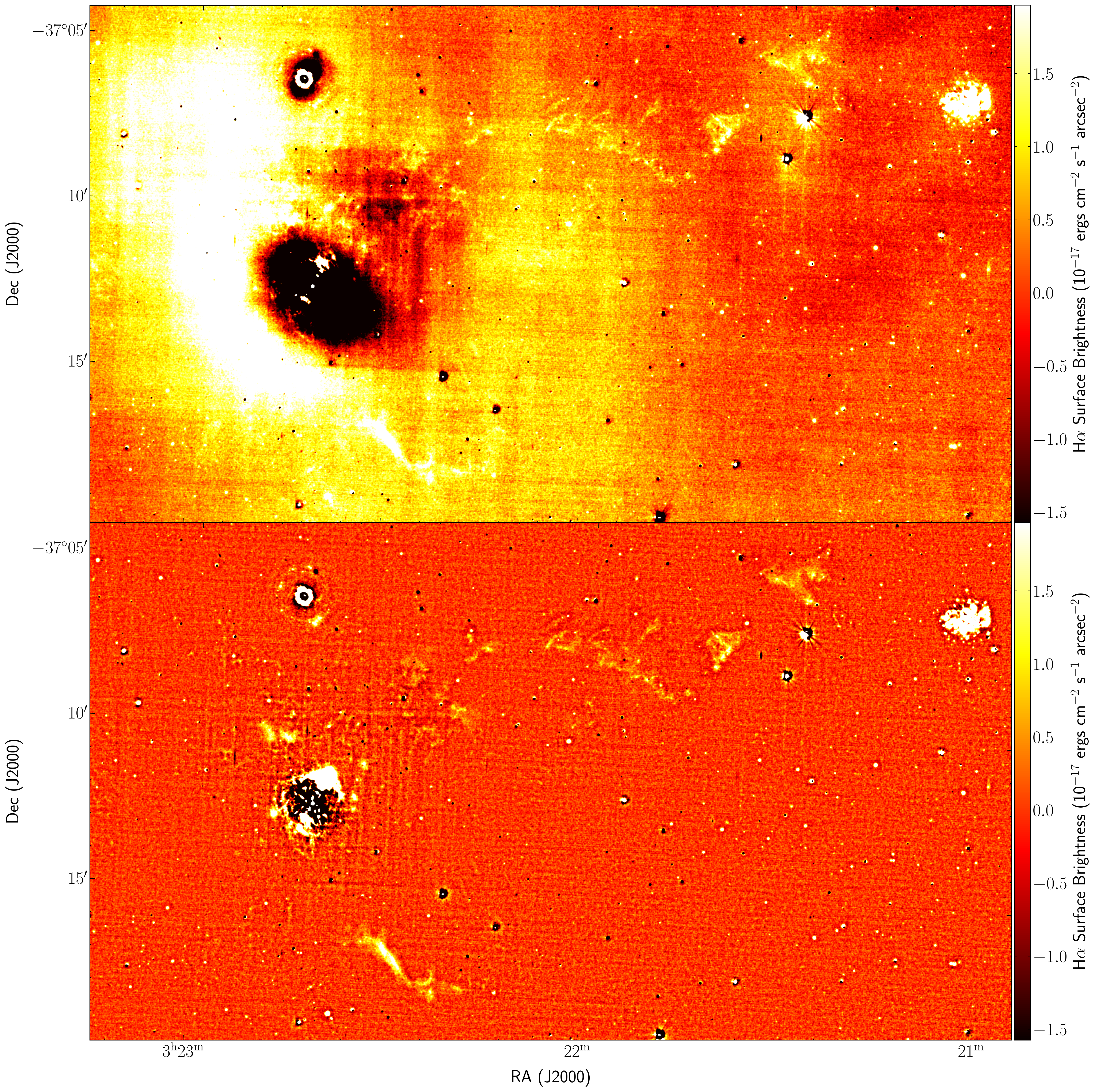}

\caption{Comparison of the original and filtered \Ha images. \emph{Top image}: \Ha image after the standard data reduction process. \emph{Bottom image}: \Ha image we present in our work that iteratively modelled and subtracted (described in section \ref{sec:Ha_obs}) the background of the original image. Both images are presented on the same scale. The original image is clearly dominated by over- and under-subtracted artefacts, while the new image has a smooth and uniform background, which retains the majority of the real \Ha emission. Some diffuse \Ha emission is lost in this process, however, the new image is a significant improvement that shows the underlying structure of the giant \Ha filaments in the IGM.}

\label{fig:Halpha_compare}
\end{figure*}

In Fig.\ref{fig:Halpha_compare}, we show the \Ha image after the standard reduction and the image we used that modelled and subtracted the background of the original image using a median filter. The giant \Ha filaments can be seen in the original image, however, it is dominated by the over- and under-subtracted artefacts and has a variable background. 

The success of our median smoothing and model background subtraction is dependent on how well the real \Ha is masked. Anything that is included in the mask, by definition, is included in the final image. This is especially challenging for diffuse \Ha emission located in areas with high background noise. The converse is also true: if spurious \Ha emission is included in the mask, it will also be in the final image.

To mitigate these issues as best as possible, we used a conservative approach to carefully mask the real \Ha that was clearly visible in the original image. It is particularly difficult to mask real \Ha emission in areas with a highly variable background and where the background is significantly under subtracted. The result is that some of the diffuse \Ha emission is lost and not reproduced in the final image. As this is an iterative process, we were able to recover \Ha emission in the most over-subtracted regions of the image. Even though we cannot conserve 100\% of the \Ha emission in this process, the purpose of this is to present the underlying structure of the new, giant \Ha filaments detected in the IGM.

%\begin{thebibliography}{References}
%\end{thebibliography}

\end{document}